\documentclass[preprint2]{aastex}    

 \usepackage{epsf}
 \usepackage{epsfig}

 \newcommand{\shapenoise}{$\sigma_{int}$}

 \newcommand{\eg}{{\it e.g.}}
 \newcommand{\cf}{{\it c.f.}}

 \newcommand{\sn}{S{\rm /}N}

 \def\deg      {{\ifmmode^\circ\else$^\circ$\fi}} 

 \def\multidrizzle{{\tt MultiDrizzle}}

\slugcomment{For the COSMOS \apj\ special edition}
 
 \shorttitle{Weak Gravitational Lensing with COSMOS.}
 \shortauthors{A.\ Leauthaud et al.}
 
\begin{document}
  

 \title{ Weak Gravitational Lensing with COSMOS: Galaxy Selection and Shape Measurements}




\author{Alexie Leauthaud\altaffilmark{1},
Richard Massey\altaffilmark{2},
Jean-Paul Kneib\altaffilmark{1,2},
Jason Rhodes\altaffilmark{2,3},\\
David E. Johnston\altaffilmark{3},
Peter Capak\altaffilmark{2},
Catherine Heymans\altaffilmark{4},
Richard S. Ellis\altaffilmark{2},\\
Anton M. Koekemoer\altaffilmark{5},
Oliver Le F\`{e}vre\altaffilmark{1},
Yannick Mellier\altaffilmark{6,7},
Alexandre R\'{e}fr\'{e}gier\altaffilmark{8},
Annie C. Robin\altaffilmark{9},
Nick Scoville\altaffilmark{2},
Lidia Tasca\altaffilmark{1},
James E. Taylor\altaffilmark{10} \&
Ludovic Van Waerbeke\altaffilmark{4}}

\email{alexie.leauthaud@oamp.fr}

\altaffiltext{1}{Laboratoire d'Astrophysique de Marseille, BP 8,
  Traverse du Siphon, 13376 Marseille Cedex 12, France.}
\altaffiltext{2}{California Institute of Technology, MC 105-24, 1200
  East California Boulevard, Pasadena, CA 91125, U.S.A.}
\altaffiltext{3}{Jet Propulsion Laboratory, Pasadena, CA 91109.}
\altaffiltext{4}{Department of Physics \& Astronomy, University of
  British Columbia, 6224 Agricultural Road, Vancouver, B.C. V6T 1Z1,
  Canada.} 
\altaffiltext{5}{Space Telescope Science Institute, 3700 San
  Martin Drive, Baltimore, MD 21218, U.S.A.}
\altaffiltext{6}{Institut d'Astrophysique de Paris, UMR7095 CNRS,
  Universit\'e Pierre~\&~Marie Curie - Paris, 98 bis bd Arago, 75014
  Paris, France.}
\altaffiltext{7}{Observatoire de Paris, LERMA, 61, avenue de
  l'Observatoire, 75014 Paris, France.}
\altaffiltext{8}{Service d'Astrophysique, CEA/Saclay, 91191
  Gif-sur-Yvette, France}
\altaffiltext{9}{Observatoire de Besan{\c c}on, BP1615, 25010 Besan{\c
    c}on Cedex, France} 
\altaffiltext{10}{Department of Physics and
  Astronomy, University of Waterloo, 200 University Avenue West,
  Waterloo, Ontario, Canada N2L 3G1}


 
  
\begin{abstract}

  With a primary goal of conducting precision weak lensing
  measurements from space, the COSMOS$^\star$ survey has imaged the
  largest contiguous area observed by the {\it Hubble Space Telescope
    (HST)} to date using the {\it Advanced Camera for Surveys
    (ACS)}. This is the first paper in a series where we describe our
  strategy for addressing the various technical challenges in the
  production of weak lensing measurements from the COSMOS data. We
  first construct a source catalog from 575 ACS/WFC tiles (1.64
  degrees$^2$) sub-sampled at a pixel scale of 0.03\arcsec. Defects
  and diffraction spikes are carefully removed, leaving a total of
  $1.2 \times 10^6$ objects to a limiting magnitude of
  $F814W=26.5$. This catalog is made publicly
  available. Multi-wavelength follow-up observations of the COSMOS
  field provide photometric redshifts for 73\% of the source galaxies
  in the lensing catalog. We analyze and discuss the COSMOS redshift
  distribution and show broad agreement with other surveys to
  $z\sim1$. Our next step is to measure the shapes of galaxies and to
  correct them for the distortion induced by the time varying ACS
  Point Spread Function and for Charge Transfer Efficiency effects.
  Simulated images are used to derive the shear susceptibility factors
  that are necessary in order to transform shape measurements into
  unbiased shear estimators. For every galaxy we derive a shape
  measurement error and utilize this quantity to extract the intrinsic
  shape noise of the galaxy sample.  Interestingly, our results
  indicate that the intrinsic shape noise varies little with either
  size, magnitude or redshift.  Representing a number density of $66$
  galaxies per arcminute$^{2}$, the final COSMOS weak lensing catalog
  contains $3.9 \times 10^5$ galaxies with accurate shape
  measurements.  The properties of the COSMOS weak lensing catalog
  described throughout this paper will provide key input numbers for
  the preparation and design of next-generation wide field space
  missions.

\end{abstract}
 

 
 


\altaffiltext{$^\star$}{Based on observations with the NASA/ESA {\em
    Hubble Space Telescope}, obtained at the Space Telescope Science
  Institute, which is operated by AURA Inc, under NASA contract NAS
  5-26555; also based on data collected at: the Subaru Telescope,
  which is operated by the National Astronomical Observatory of Japan;
  the European Southern Observatory, Chile; Kitt Peak National
  Observatory, Cerro Tololo Inter-American Observatory, and the
  National Optical Astronomy Observatory, which are operated by the
  Association of Universities for Research in Astronomy, Inc. (AURA)
  under cooperative agreement with the National Science Foundation;
  the National Radio Astronomy Observatory which is a facility of the
  National Science Foundation operated under cooperative agreement by
  Associated Universities, Inc ; and the Canada-France-Hawaii
  Telescope operated by the National Research Council of Canada, the
  Centre National de la Recherche Scientifique de France and the
  University of Hawaii.}


\section{Introduction}


As we look towards distant galaxies, fluctuations in the intervening
mass distribution cause a slight, coherent distortion of their
intrinsic shapes. This effect, known as weak gravitational lensing,
has been used for more than a decade to probe the cosmography and the
growth of structure \citep[for a review,
see][]{Bartelmann:2001}. Although technically challenging because the
weak lensing signal is minuscule and buried in a considerable amount
of noise, this field has shown substantial progress due to the advent
of high resolution space based imaging, the proliferation of
wide-field multi-color surveys, and a determined effort to improve
image analysis methods and minimize systematic errors through the
Shear TEsting Program \citep[STEP;][]{Heymans:2006a,Massey:2006}.

Historically first observed only around cluster cores
\citep[][]{Tyson:1990}, weak lensing has emerged as a versatile and
effective technique to probe the mass distribution of clusters
\citep[e.g.,][]{Kneib:2003}, to measure the clustering of dark matter
around galaxies ensembles
\citep[e.g.,][]{Natarajan:1998,Hoekstra:2004, Sheldon:2004, Mandelbaum:2005},
and to put constraints on the matter density parameter $\Omega _m$ and
the amplitude of the matter power spectrum $\sigma _8$
\citep[e.g.,][]{Hoekstra:2006,Semboloni:2006,Schrabback:2006,Hetterscheidt:2006,Jarvis:2006}. Most
applications currently only use the first order deformation induced by
the mass distribution, but novel techniques are under development to
take into account second order deformations \citep[also called
\textit{flexion};][]{Goldberg:2005,Bacon:2006,Okura:2006,Goldberg:2006}
and may prove to be more efficient probes of compact structures such
as galaxies and groups of galaxies. Weak lensing measurements are
particularly powerful when combined with the knowledge of the three
dimensional galaxy distribution. Sophisticated lensing 'tomography'
techniques that utilize redshifts to analyze the 3D shear field are a
sensitive probe of the growth of structure and the equation of state
of dark energy
\citep[][]{Jain:2003,Bernstein:2004,Bacon:2005}. Applied in many
different ways, weak lensing techniques unravel the mass distribution
of structures and their evolution in the universe.


High quality measurements of weak shear depend on the accurate
determination of the shapes and redshifts of distant, faint galaxies.
The COSMOS program has imaged the largest contiguous area (1.64
degrees$^2$) with the {\it Hubble Space Telescope (HST)} to date using
the {\it Advanced Camera for Surveys (ACS)} {\it Wide Field Channel
  (WFC)}. The Full Width Half Maximum (FWHM) of the Point Spread
Function (PSF) of the ACS/WFC is $0.12\arcsec$ at the
detector\footnote{Before convolution with the detector pixels, the
  intrinsic width of the F814W PSF is 0.085\arcsec}, yielding a much
better resolution of small galaxies than ground-based surveys, which
are typically limited by the seeing to a PSF of FWHM $ \sim
1\arcsec$. Shape measurements also benefit from ACS/WFC imaging
compared to ground-based observations because smaller corrections are
required for the PSF and the shear measurements are less diluted by
PSF smearing. The imaging quality and unprecedented area of the COSMOS
ACS/WFC data combined with extensive follow-up observations at other
wavelengths \citep{Scoville:2007,Koekemoer:2007} to provide accurate
photometric redshifts \citep{Mobasher:2007}, make COSMOS a unique data
set for weak lensing studies.

To exploit the weak lensing potential of the COSMOS ACS/WFC data, a
carefully designed catalog of resolved galaxies with shape
measurements must be extracted from the imaging data. The challenges
and requirements of such a catalog are the following.  First, the
large survey size makes a robust automation of catalog generation
essential. Second, the lensing sensitivity increases with the number
density of resolved faint galaxies. Thus, it is important to detect
all galaxies to faint magnitudes while taking care to minimize
spurious detections that will add noise to weak lensing
measurements. Third, the high spatial resolution of the ACS/WFC allows
for an excellent separation of close pairs.  Given accurate deblending
and a high number density of galaxies, one can expect to measure shear
statistically on sub-arcminute scales, where baryonic physics may
begin to influence the dark matter distribution.  As the
COSMOS-ACS/WFC data set is likely to be the only large space-based
lensing survey until the launch of next-generation wide field space
missions such as \textsc{snap} \footnote{http://snap.lbl.gov} or
\textsc{dune} \citep{Refregier:2006}, the knowledge acquired through
the COSMOS data will be unique and of crucial importance for the
preparation and design of these future missions.

This is the first paper of a series describing the galaxy selection,
the galaxy shape measurement, the lensing analysis, and the
cosmological interpretation of COSMOS data. Details regarding PSF
corrections as well as tests for systematic effects are presented in
the second paper of this series \citep{Rhodes:2007}.  A three
dimensional cosmic shear analysis is presented in the third paper of
this series \citep{Massey:2007}. Finally, a fourth paper presents high
resolution dark matter mass maps of the COSMOS field
\citep{Massey:2007a}.

In this paper, we describe our methods for constructing a galaxy
catalog from the ACS/WFC data, to be used in subsequent weak lensing
work with COSMOS. Our goal is to produce a catalog of galaxies with
photometric redshifts and PSF-corrected shape measurements, free of
contaminating stars, cosmic rays, diffraction spikes, and other
artifacts.  The paper is organized as follows. In $\S$\ref{data} we
present the data. In $\S$\ref{catalog} we describe the pipeline that
locates and measures the properties of all detected objects. In
$\S$\ref{quality}, we assess the quality of the data and analyze the
COSMOS redshift distribution. In $\S$\ref{finalcat} and $\S$\ref{cuts}
we present the PSF and CTE correction schemes, the shape and shear
measurement methods, and our selection criteria for the final lensing
catalog. In $\S$\ref{shapenoise} we extract the intrinsic shape noise
of the galaxy sample as a function of redshift and discuss the
implications for future weak lensing surveys. Where necessary, we
assume a standard cosmological model with $\Omega_{\rm M}=0.3$,
$\Omega_\Lambda=0.7$, $H_0=100 h$ km~s$^{-1}$~Mpc$^{-1}$ and $h =
0.7$.


\section{The COSMOS ACS Data}\label{data}

\begin{figure}[tb]
\epsscale{1.0}
\plotone{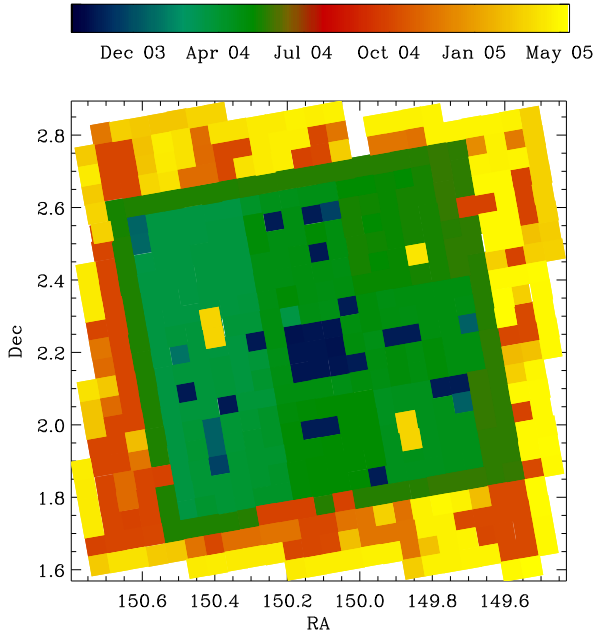}
\caption{  Date of observation for each of the survey pointings. The PSF of the
  ACS/WFC varies on timescales that are much shorter than the period over which
  COSMOS was observed.}
\label{fig:date_of_observation}
\end{figure}

\begin{figure*}[p]
\psfig{file=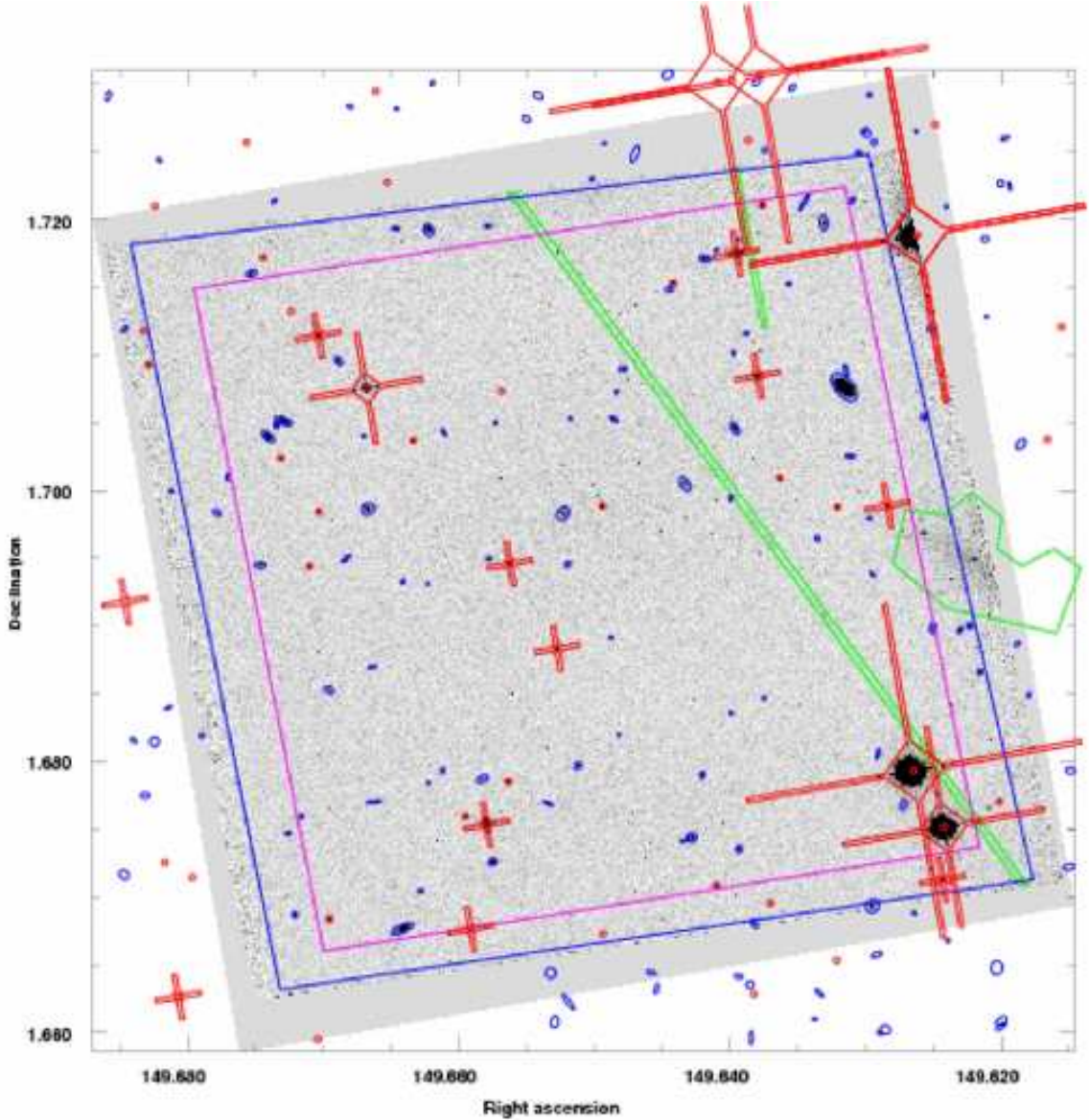,width=\textwidth}
\caption{ COSMOS pointing acs\_I\_095836+0141\_unrot\_sci\_12.fits
  with bright detections, masking and edge definition
  indicated. Adjacent images have a sizable overlap (shown here by the
  smaller magenta box) which allows us to discard detections on the
  boundaries of each tile (defined by the larger blue box) without
  losing any objects in the final concatenated catalog. The automated
  masking of the diffraction spikes around bright stars ($F814 < 23$)
  is pictured here by the red polygons. The basic shape of the star
  masks is predefined and then scaled with the magnitude of the star.
  The green rectangles correspond to the manual masking of asteroid
  trails and various other image defects. Toward the right edge of
  this image, the dwarf galaxy L1-099 \citep[][]{Impey:1996} is
  identified and flagged in a special category. Bright galaxies with
  $F814 < 23$ are depicted by blue ellipses and bright stars with
  $F814 < 23$ by red circles.}
\label{image}
\end{figure*}

The COSMOS HST ACS field \citep{Scoville:2007,Koekemoer:2007} is a
contiguous 1.64 degrees$^2$, centered at 10:00:28.6, +02:12:21.0
(J2000).  Between October 2003 and June 2005 (HST cycles 12 and 13),
the region was completely tiled by 575 adjacent and slightly
overlapping pointings of the ACS/WFC (see
Figure~\ref{fig:date_of_observation}). Images were taken through the
wide F814W filter (``Broad I'').  The camera has a $203\arcsec\times
203\arcsec$ field of view, covered by two 4096x2048 CCD chips with a
native pixel scale of $0.05\arcsec$ \citep{Ford:2003}. The median
exposure depth across the field is 2028 seconds (one HST orbit). At
each pointing, four 507 second exposures were taken, each dithered by
0.25$\arcsec$ in the $x$ direction and 3.08$\arcsec$ in the $y$
direction from the previous position. This strategy ensures that the
3$\arcsec$ gap between the two chips is covered by at least three
exposures and facilitates the removal of cosmic rays. Pointings were
taken with two approximately $180\arcdeg$ opposed orientation angles
({\ttfamily PA\_V3}$=100 \pm 10\arcdeg$ and $290\pm 10\arcdeg$). In
this paper we use the ``unrotated'' images (as opposed to North up) to
avoid rotating the original frame of the PSF. By keeping the images in
the default unrotated detector frame, they can be stacked to map out
the observed PSF patterns. For similar reasons, we perform detection
in individual ACS/WFC tiles instead of on a larger mosaic (where the
orientation of the PSF frame would be unknown). Figure \ref{image}
shows a COSMOS/ACS pointing with the bright detections ($F814W < 23$)
and the masking of stars, asteroid trails and image defects (see
$\S$\ref{clean}).

To build our catalog, we use version 1.3 of the ``unrotated'' ACS/WFC
data which has been specially reduced for lensing purposes
\citep[see][for technical details]{Koekemoer:2007}. Image
registration, geometric distortion, sky subtraction, cosmic ray
rejection and the final combination of the dithered images were
performed by the \multidrizzle\ algorithm \citep{Koekemoer:2002}.  As
described in \citep{Rhodes:2007}, the \multidrizzle\ parameters have
been chosen for precise galaxy shape measurement in the co-added
images. In particular, a finer pixel scale of $0.03\arcsec/$pix was
used for the final co-added images (7000x7000 pixels), even though
this implies more strongly correlated pixel noise (see
\S\ref{noise}). Hereafter, when we refer to pixels, we will assume a
pixel scale of $0.03\arcsec/$pix. Pixelization acts as a convolution
followed by a re-sampling and, although current shear measurement
methods can successfully correct for convolution, the formalism to
properly treat re-sampling is still under development for the next
generation of methods.  Again following the recommendations of
\citet{Rhodes:2007}, a Gaussian and isotropic multidrizzle\
convolution kernel was used, with {\ttfamily scale}=0.6 and {\ttfamily
  pixfrac}=0.8, small enough to avoid smearing the object
unnecessarily while large enough to guarantee that the convolution
dominates the re-sampling. This process is then properly corrected by
existing shear measurement methods.


The ACS/WFC CCDs also suffer from imperfect charge transfer efficiency
(CTE) during read-out. As charges are transferred during the read-out
process, a certain fraction are retained by charge traps (created by
cosmic ray hits) in the pixels. This causes flux to be trailed behind
objects as the traps gradually release their charge, spuriously
elongating them in a coherent direction that mimics a lensing
signal. Since this effect is produced by a fixed number of charge
traps within the CCD substrate, it affects faint sources (with a
larger fraction of their flux being trailed) more than bright
ones. This is an insidious effect that mimics an increasing shear
signal as a function of redshift, and prevents the traditional way of
dealing with the calibration of faint galaxies in a lensing analysis
by looking at bright stars. Ideally this effect would be corrected for
on a pixel-by-pixel basis in the raw images unfortunately our current
physical understanding of this effect is insufficient and a more
indepth analysis in still underway. The CTE effect can be quantified
sufficiently well however that, in a first step, we can adopt a
post-processing correction scheme based on an object's position, flux,
and date of observation. Further details regarding this model can be
found in \citet{Rhodes:2007}.

\section{The COSMOS ACS Galaxy Catalog}\label{catalog}

In this section, we discuss the construction of the COSMOS ACS/WFC
source catalog. This catalog is carefully cleaned of defects and
artifacts and is made publicly available through the Infrared Science
Archive (IRSA) database\footnotemark[1].

\footnotetext[1]{ http://irsa.ipac.caltech.edu/Missions/cosmos.html}

\subsection{Detection Strategy}\label{strategy}


We use Version 2.4.3 of the SExtractor photometry package
\citep{Bertin:1996} to extract a source catalog of positions and
various photometric parameters. In the construction of this catalog,
our main concern is to pick out the small faint objects that contain
most of the lensing signal. The detection strategy that we therefore
adopt is to configure SExtractor with very low thresholds (even if
this leads to more false detections in the catalog) and to control our
sample selection via subsequent ``lensing cuts'' (see Sect
\ref{cuts}).  We hope to thus reduce unknown selection biases
introduced by the SExtractor detection algorithm. When configured with
low detection thresholds however, SExtractor also inevitably a)
overdeblends low surface brightness spirals and patchy irregulars, b)
deblends the outer features of bright galaxies, c) detects spurious
objects in the scattered light around bright objects and d)
underdeblends close pairs.

The correct detection of close pairs enables lensing measurements on
very small scales. However, overdeblending and spurious detections
adds noise to these measurements. In particular, false detections
around bright objects can have quite high signal-to-noise ($\sn$)
values and are not trivial to remove with lensing cuts ( see
$\S$\ref{cuts}). The method presented here is a partial solution to b)
c) and d). The overdeblending of low surface brightness and patchy
galaxies remains a difficult problem, however, especially for high
resolution imaging, and calls for an improvement of existing detection
algorithms.  With the advent of high resolution multi-wavelength
surveys, a possible solution forward would be to incorporate color and
morphological information into the detection process
\citep[e.g.,][]{Lupton:2001}.

While this overdeblending problem persists in our catalog, it affects
less than $1\%$ of the objects. This problem is furthermore mitigated by
the centroiding process during the shape measurement stage. Indeed,
objects for which the centroid algorithm fails to converge, which will
often be the case for overdeblended features, are discarded from the
catalog. To remedy the remaining problems b) c) and d), we adopt and
improve the method (known as the \textit{`Hot-Cold`} method) employed
in \citet{Rix:2004}.  In this method, we run SExtractor twice, once
with a configuration optimized for the detection of only the brightest
objects (``cold'' step) and then again with a configuration optimized
for the faint objects (``hot'' step).  This double extraction helps
improve the detection of close pairs. The two samples are then merged
together to form the final catalog and masks are created around the
bright detections minimizing the effects of c) and d).

For the ``hot'' and ``cold'' steps we vary four main parameters to
optimize the detection: 1) \textsc{detect\_threshold}, the minimum
signal-to-noise per pixel above the background level, 2)
\textsc{min\_area}, the number of contiguous pixels exceeding this
threshold, 3) \textsc{back\_size}, the mesh size of the background
map, 4) \textsc{deblend\_nthres} and \textsc{deblend\_mincont}, the
parameters regulating deblending. In both cases, the data are filtered
prior to detection by a 5 pixel (0.15\arcsec) Gaussian filtering
kernel. Our choice of parameters for both steps is provided in Table
\ref{se}.



\begin{deluxetable}{lcc}
\tabletypesize{\scriptsize}
\tablecolumns{3}
\tablecaption{SExtractor configuration parameters\label{se}}
\tablewidth{0pt}
\tablehead{
\colhead{Parameter} & \colhead{Bright objects}  & \colhead{Faint objects}  } 
\startdata

DETECT\_MINAREA & 140 & 18   \\ 
DETECT\_ THRESH  & 2.2\tablenotemark{*} & 1.0   \\
DEBLEND\_ NTHRESH & 64 & 64  \\
DEBLEND\_MINCONT & 0.04 & 0.065  \\
CLEAN\_PARAM & 1.0 & 1.0  \\
BACK\_SIZE & 400 & 100 \\
BACK\_FILTERSIZE & 5 & 3 \\
BACKPHOTO\_TYPE & local & local \\
BACKPHOTO\_THICK & 200 & 200 \\

\enddata
\tablenotetext{*}{Because of correlated noise (see $\S$
  \ref{noise}), the effective threshold levels are \textsc{DETECT\_THRESH} $\sim 1.25$ for
  bright objects and \textsc{DETECT\_THRESH} $\sim 0.57$ for faint objects.}
\end{deluxetable}

The two-step method also allows one to adjust the estimation of the
background map according to the typical size of objects one expects to
detect, improving detections with SExtractor. The background map is
constructed by computing an estimator for the local background on a
grid of mesh size \textsc{back\_size}. We adjust \textsc{back\_size}
so as to capture the small scale variations of the background noise
while keeping it large enough not to be affected by the presence of
objects.

For each exposure, a weight map is produced by \multidrizzle\
describing the combined noise properties of the read-out, the dark
current and the sky background (\citet{Koekemoer:2007}). These maps
describe the noise intensity at each pixel and are used to account for
the spatial-dependent noise pattern in the co-added image with the
SExtractor \textsc{weight\_image} option set to \textsc{weight map}.

Each ACS/WFC pointing consists of four slightly offset, dithered
exposures making cosmic ray rejection more difficult and detection
more unreliable on the boundaries of each tile where there are fewer
than four input exposures. Because adjacent images overlap
sufficiently, we can trim the edges of the images without actually
removing data (see Figure \ref{image}).

\subsection{Bright Object Detection}

\begin{figure}[htb] 
\plotone{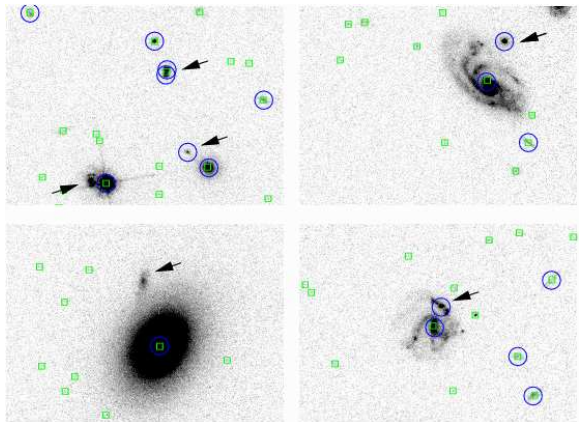}
\caption{ This figure illustrates the difficulty of correctly
  deblending close pairs while keeping patchy spirals with strong star
  forming regions intact. Squares indicate detections from the faint
  step and circles indicate detections from the bright step. The top
  two panels show three objects that are not detected or incorrectly
  deblended by the faint step but that are picked by two step
  method. The bottom panel and the arrow towards the star in the upper
  left panel show that even with this method, a perfect configuration
  is still difficult to reach.}
\label{hotcold}
\end{figure}

In the first step we detect only the brightest and largest objects in
the image, with 140 or more contiguous pixels (corresponding to a
diameter of 0.4\arcsec\ for a circular object) rising more than 2.2
sigma per pixel above the background level.  The \textsc{back\_size}
parameter is set to 12\arcsec, or 30 times the diameter of the
smallest objects detected. The detection threshold and the deblending
parameters \textsc{deblend\_nthres \& deblend\_mincont} are calibrated
heuristically on several images to separate close pairs as much as
possible without deblending patchy, extended spiral galaxies. Because
faint objects are captured in a second run, we are free to choose the
value of the detection threshold during this step. We found that this
flexibility greatly helped to calibrate the parameters that optimize
the deblending. Indeed, if \textsc{detect\_threshold} is set to a low
value (for example, to detect faint object), close pairs will be
detected as a single object and are difficult to deblend. Figure
\ref{hotcold} illustrates the improvements of this two-step method
compared to a single-step method. During this first step, all pixels
associated with a detection are recorded by SExtractor in an image
called a ``segmentation map''. These segmentation maps are used at a
later stage to merge the bright and the faint catalogs (see
$\S$\ref{merge}). This first catalog of bright objects is referred to
as $\mathcal{C}_{cold}$.

\subsection{Faint Object Detection}

In the second step, we configure SExtractor to pick up the small,
faint objects, taking care to choose the detection parameters to be
less conservative than any subsequent lensing cuts (see
$\S$\ref{cuts}).  \textsc{min\_area} is set to 18 pixels
(corresponding to a diameter of 1.2 times the FWHM of the PSF) and the
detection threshold is one sigma above the background level.  As
objects detected at this step are smaller, the background estimation
can be improved by refining the mesh size of the background map and
setting \textsc{back\_size} to 100 pixels, or 20 times the diameter of
the smallest objects detected.  This second catalog is referred to
hereafter as $\mathcal{C}_{hot}$.

\subsection{Merging the Two Samples} \label{merge}

The final catalog is obtained by merging the detections from
$\mathcal{C}_{cold}$ and $\mathcal{C}_{hot}$, keeping all objects in
$\mathcal{C}_{cold}$ and only the objects from $\mathcal{C}_{hot}$ not
detected in $\mathcal{C}_{cold}$. To determine which objects to
discard from $\mathcal{C}_{hot}$, we use the segmentation maps created
during the bright detection step. To begin with, we enlarge the
flagged areas in these segmentation maps by approximately 20 pixels
(0.6$\arcsec$). We then discard all objects from $\mathcal{C}_{hot}$
for which the central pixel lies within a flagged area of these
maps. Thus, we remove duplicate detections and create a mask around
all bright objects, immediately cleaning the catalog of a certain
number of spurious detections. By a visual inspection of the data, we
estimate that this method solves about half of the deblending problems
that we observe (excluding the low surface brightness galaxies). The
final catalog of raw SExtractor detections $\mathcal{C}_1$ contains
1.8 million objects in total (see Table \ref{final_table}).



\begin{deluxetable}{lcc}
\tabletypesize{\scriptsize}
\tablecaption{Sumary of the construction of the ACS lensing catalog}
\tablewidth{0pt}
\tablecolumns{3}
\tablehead{
\colhead{Catalog of raw SExtractor detections: $\mathcal{C}_1$} & \colhead{Number}  & \colhead{Percent of $\mathcal{C}_1$}
}
\startdata
Total number of objects in $\mathcal{C}_1$      & 1.8 $\times 10^6$ &  \\  
Number of Hot (faint) detections      & 1.6$\times 10^6$ & 88\%  \\                     
Number of Cold (bright) detections      & 2.2$\times 10^5$ & 12\% \\                    

\cline{1-3} \\
\multicolumn{1}{c}{Details of the cleaning process} & Number & Percent of $\mathcal{C}_1$\\
\cline{1-3} \\

Number of objects within the noisy border of a tile & 3.2$\times 10^5$ & 17\%\\  
Number of Hot detections with central pixel in Cold segmentation map      & 2.0$\times 10^5$ & 11\%\\ 
Number of objects within automatically defined star masks  & 2.4$\times 10^4$ & 1\%\\  
Number of objects within manually defined masks  & 4.1$\times 10^4$ & 2\%\\ 
Number of objects detected more than once in adjacent tiles  & 6.6$\times 10^4$ & 4\%\\ 

\cline{1-3} \\
\multicolumn{1}{c}{Catalog cleaned of image defects: $\mathcal{C}_2$} & Number & Percent of $\mathcal{C}_2$\\
\cline{1-3} \\

Total number of objects in $\mathcal{C}_2$  & 1.2$\times 10^6$ & \\ 
Number of galaxies ($ID=1$)  & 1.1$\times 10^6$ & 96 \%\\ 
Number of point sources ($ID=2$)  & 2.8$\times 10^4$ & 2 \%\\ 
Number of fake detections ($ID=3$)  & 1.7 $\times 10^4$ & 2\%\\ 

\cline{1-3} \\
\multicolumn{1}{c}{ACS galaxies from $\mathcal{C}_2$ with $F814W_{AB} < 26.5$: $\mathcal{C}_3$} & Number & Percent of $\mathcal{C}_3$\\
\cline{1-3} \\

Total number of objects in $\mathcal{C}_3$ & 7.0$\times 10^5$ & \\ 
Number of galaxies with a counterpart in the photometric catalog  & 6.0$\times 10^5$ & 85\%\\ 
Number of galaxies that have been matched but that are in ground based masks  & 8.3$\times 10^4$ & 12\%\\ 
Number of galaxies for which the redshift code did not converge  & 1.1$\times 10^4$ & 1.7\% \\  
Total number of galaxies with accurate photometric redshifts, $\mathcal{C}_3$  & 5.0$\times 10^5$ & 71\%\\

\cline{1-3} \\
\multicolumn{1}{c}{The final COSMOS ACS lensing catalog: $\mathcal{C}_4$} & Number & Galaxy number density\\
\cline{1-3} \\

Total number of galaxies in final lensing catalog  & 3.9$\times 10^5$ & 66 arcmin$^2$\\
Number of galaxies with accurate photometric redshifts  & 2.8$\times 10^5$ & 48 arcmin$^2$\\

\enddata
\label{final_table}
\end{deluxetable}


\subsection{Cleaning the Catalog}\label{clean}

Great care was taken to mask unreliable regions within images and to
remove false detections from the catalog, especially those that can
mimic a lensing signal. As illustrated in Figure~\ref{image}, an
automatic algorithm was developed to define polygonal shaped masks
around stars with $F814W< 19$ (the limit at which stars saturate in
the COSMOS images), with a size scaled by the magnitude of the star.
Objects near bright stars or saturated pixels were masked to avoid
shape biases due to any background gradient. All the images were then
visually inspected. In a few cases the automatic algorithm failed
(very saturated stars for which the centroid of the star is widely
offset) and the stellar masks were corrected by hand. Other
contaminated regions of the images were also masked out, including
reflection ghosts, asteroids, and satellite trails. Astronomical
sources such as HII regions around bright galaxies, stellar clusters,
and nearby dwarf galaxies, were also flagged and removed from the
lensing catalog.

Objects with double entries in the catalog (from the overlap between
adjacent images) are identified and the counterpart with the highest
SExtractor flag (indicating a poor detection) is discarded, leaving a
catalog of unique objects.  However, the duplicated objects from
overlapping regions are a valuable asset for consistency checks and
are used, for example, to check the galaxy shape measurement error
(see $\S$\ref{shapenoise} and Figure \ref{sigma_mag}).

The final clean catalog ($\mathcal{C}_2$) is free of spurious or
duplicate detections and contains 1.2 million sources in total (see
Table \ref{final_table}).

\subsection{Star-Galaxy Classification}

The correct identification of stars has two implications for the
lensing analysis. First, bright stars are useful for PSF modelling and
second, stars must be correctly identified in order to apply our
automatic masking algorithm of diffraction spikes. A robust
star-galaxy classification is thus necessary.

\begin{figure}[tb] 
\epsscale{1.1}
\plotone{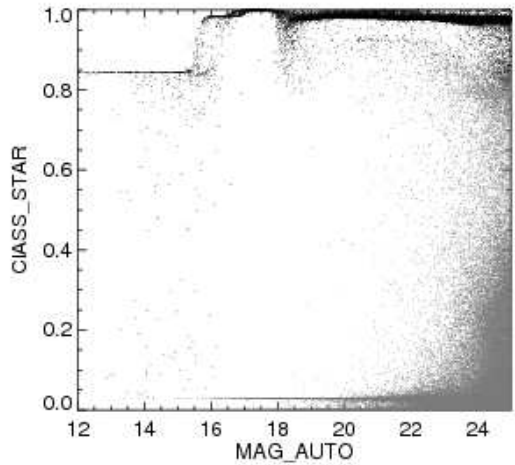}
\caption{ The SExtractor stellar index (\textsc{class\_star}) for our point
  source selection based on the peak surface brightness of objects.
  Grey points show the corresponding galaxy sample. A point source
  selection of the form $\textsc{class\_star} > 0.8$ for example, will
  miss a certain number of bright stars ($F814W < 17$) and will
  grossly misclassify compact galaxies at $22<F814W<25$.}
\label{classstar}
\end{figure}

{\sc SExtractor} produces a continuous stellar classification index
parameter ranging from 0 (extended sources) to 1 (point sources). This
index has two drawbacks: first the definition of the dividing line is
ambiguous and second, the neural-network classifier used by {\sc
  SExtractor} is trained with ground-based images and is therefore
only valid for a sample of profiles similar to the original training
set. With space-based images, this index becomes difficult to
interpret, as illustrated in Figure \ref{classstar} which depicts our
star selection (described below) within the
\textsc{class\_star}/\textsc{mag\_auto} plane.

\begin{figure}[tb] 
\epsscale{1.1}
\plotone{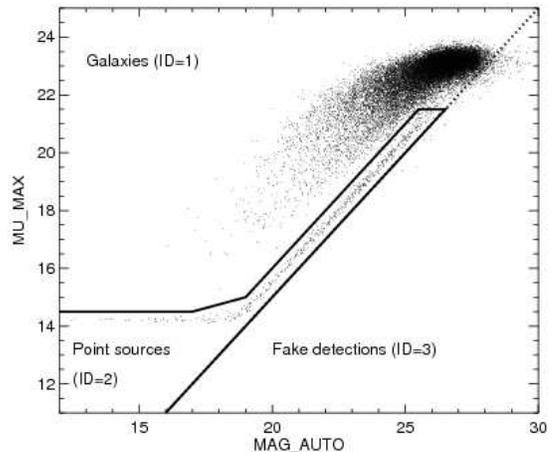}
\caption{Classification of point sources, galaxies and artifacts within the
  \textsc{mu\_max}/ \textsc{mag\_auto} plane. Point sources follow the
  PSF and are delimited by the solid region. Objects that are more
  sharply peaked than the PSF are contained in the dashed region and
  are considered to be artifacts.  For clarity, only a 2\% random
  selection of all objects are included in this plot.}
\label{mutype}
\end{figure}

We therefore test two alternative methods to classify point sources
and galaxies, one based on the SExtractor parameter \textsc{mu\_max}
(peak surface brightness above the background level) and the other
based on the half-light radius, \textsc{Rhl} \citep[e.g.,
][]{Peterson:1979,Bardeau:2005}. Both methods are motivated by
the fact that the light distribution of a point source scales with
magnitude. Point sources therefore occupy a well-defined locus in a
\textsc{mu\_max}/\textsc{mag\_auto} or a
\textsc{Rhl}/\textsc{mag\_auto} plane. Figure \ref{mutype} shows how
we can use this property to define stars (ID=2) and galaxies (ID=1)
reliably up to a magnitude of $F814W \sim 25$. At fainter levels, the
classification begins to break down and the point sources become
indistinguishable from the small galaxies. We find that the two
methods agree very well, within 1\% at magnitudes less than $F814W=24$
and within 2\% at magnitudes less than $F814W=25$. The small
difference arises mainly from a misclassification of objects by the
\textsc{Rhl} method because of the presence of a close pair that
distorts the estimation of \textsc{Rhl}.  Overall, the
\textsc{mu\_max} method proved to be more robust and has the advantage
of a tighter correlation of the stellar locus and a clear break
indicating the magnitude at which the stars saturate ($
\textsc{mag\_auto} \sim 19$). Moreover, with this method, a surface
brightness cut at the faint end of the stellar sample is trivial to
implement (at faint magnitudes, the catalog is surface brightness
limited). The performance of this star galaxy separation scheme will
be analyzed in more detail in $\S$ \ref{galcounts}.

Using the \textsc{mu\_max} method, we also define a set of objects
that are more sharply peaked than the PSF, which is obviously
non-physical. A visual inspection finds that these objects are mainly
artifacts, hot pixels, and residual cosmic rays. We flag these
spurious objects in our catalog (ID=3) and remove them for the lensing
analysis.

Averaging over the COSMOS field, we find $\sim15$ stars per pointing
with $ 19 \leq \textsc{mag\_auto} \leq 23$. This is an insufficient
number to model the PSF in individual images using standard
interpolation techniques. However, it is a sufficient number to
identify the PSF pattern of each exposure given a finite set of
recurring patterns (see $\S$ \ref{psf} and \citet{Rhodes:2007} for
further details).
 

\subsection{Photometric Redshifts}\label{photoz}

In addition to the ACS/WFC ($F814W$) imaging, the COSMOS field has
been imaged with Subaru Suprime-Cam ($B_j$,$V_j$,$g+$,$r+$,$i+$,
$z+$,$NB816$), the Canada-French Hawaii Telescope (CFHT) ($u*,i*$) and
the KPNO/CTIO ($Ks$). Details of the ground-based observations and the
data reduction are presented in \citet{Capak:2007} and
\citet{Taniguchi:2007}. Other observations were taken in the UV with
{\sl GALEX}, in the X-ray with {\sl XMM-Newton} and in the radio with
VLA, CSO and IRAM.  Yet more observations are underway including
intermediate and narrow-band imaging with Subaru Suprime-Cam, deep
Infrared imaging covering 1.0-2.2 microns (WIRCam/CFHT, WFCAM/UKIRT
and ULBcam/UH2.2), and observations with space-based facilities
including {\sl Chandra} and {\sl Spitzer} Space Telescopes. This
extensive multi-wavelength data-set is a key component to COSMOS weak
lensing measurements because it allows us to accurately measure the
COSMOS redshift distribution, to separate foreground and background
structures, and to remove contamination from intrinsic galaxy
alignments \citep{Heymans:2003} and shear-ellipticity correlations
\citep{King:2005}.

Photometric redshifts were determined by the COSMOS photometric
redshift code with a Bayesian prior based on luminosity functions and
allowing for internal extinction \citep{Mobasher:2007}. For each
galaxy, the entire probability distribution $P(z)$, the most likely
redshift and a confidence level for that redshift is calculated. The
knowledge of the full $P(z)$ allows us to apply weight to galaxies in
weak lensing measurements according to the uncertainty in their
measured redshift.  The accuracy of the photometric redshifts are
estimated based on extensive simulations and by comparison with a
sample of 958 galaxies with spectroscopic redshifts measured by the
ESO/VIMOS instrument as part as the zCOSMOS program
\citep{Lilly:2007}. Both the luminosity prior and the extinction
corrections have been shown to improve the accuracy of the photometric
redshifts when compared to the spectroscopic sample.  The accuracy of
the current photometric redshifts down to $F814W\sim 22.5$ is:

\begin{equation}\label{photoz_err}
\sigma_{\Delta z}/ (1+z_s) = 0.031
\end{equation}

\noindent with $\eta = 1.0 \%$ of catastrophic errors, defined as
$\Delta z /(1+z_s) >0.15$. This relation scales with magnitude in a
similar fashion to \cite{Wolf:2004}. The accuracy of the photometric
redshifts will continue to improve as more data become available (in
particular the deeper $u*$, $J$, $K$, Spitzer and Subaru narrow band
data).

\begin{figure}[htb] 
\epsscale{0.92}
\plotone{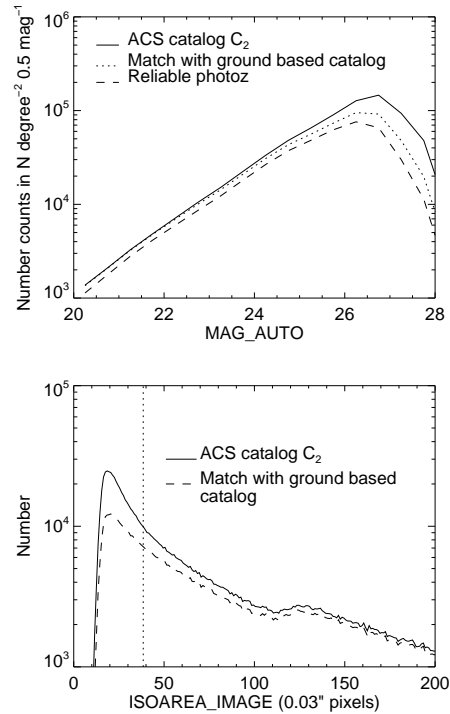}
\caption{ In the upper panel, the number counts of galaxies that have
  been correctly matched to the ground based catalog (dotted line) are
  compared to the total number counts (solid line). The dashed line
  indicates galaxies for which we consider the photometric redshift to
  be reliable. The difference between these two curves is primarily
  due to larger masked areas in the ground-based data than in the ACS
  imaging. In the lower panel, the sizes of galaxies within the ACS
  catalog (solid line) are compared to the sizes of those that have
  been matched with the ground based catalog (dashed line). The
  objects for which we do not have multi-band photometry are small
  galaxies that are detected with ACS but not with ground based
  imaging (seeing $\sim 1\arcsec$). The vertical dashed line shows the
  approximate size cut that we make in the final lensing catalog.}
\label{photozcounts}
\end{figure}

The COSMOS optical and near infrared catalog \citep{Capak:2007}
provides multi-band photometry for 89\% of the COSMOS ACS/WFC
galaxies. As demonstrated by Figure \ref{photozcounts}, the remaining
11\% of the galaxies for which we lack multi-wavelength information
(and therefore photometric redshifts), are the small galaxies that
cannot be detected by the ground based imaging. In effect, although
the SUBARU data is deeper for sources 1\arcsec\ in diameter, the
ACS/WFC imaging will do a better job at detecting anything
smaller. Because we apply a size cut to the the final lensing catalog
however (see $\S$ \ref{cuts}), many of these small galaxies will be
discarded from the final analysis. In total, after removing the
galaxies with unreliable multi-band photometry (because they are
masked out in the ground-based data) as well as those for which the
photometric redshift code failed to converge, 76\% of the galaxies in
the COSMOS ACS/WFC lensing catalog ($F814W < 26$) have photometric
redshifts (See Figure \ref{photozcounts}). For further discussions on
the photometric redshifts and the COSMOS redshift distribution, see
$\S$ \ref{redshift distribution}.

 
\subsection{Noise Properties}\label{noise}

The drizzling process introduces pattern-dependent correlations
between neighboring pixels and can artificially reduce the noise
levels in co-added images. Noise and error estimates derived from
drizzled images will thus tend to underestimate the true noise levels
of the image. One should in theory take into account the exact
covariance matrix of the noise in order to derive error estimates for
drizzled images. For our purposes, however, a simple scaling of the
noise level is each pixel by the same constant factor is
sufficient. The scaling factor that we adopt, $F_A$, has been derived
for \multidrizzle d images by \citet{Casertano:2000}. In principle
$F_A$ is size dependent but converges rapidly with increasing size
toward an asymptotic value given by:

\begin{equation}
 \sqrt{F_A} = \left\{
              \begin{array}{ll}
                   (s/p) (1 - s/(3p)) & (s < p)\\
                   (1 - p/ (3s)) & (p < s)
              \end{array}
       \right. 
\end{equation}

\noindent where $p$ and $s$ are respectively the {\ttfamily pixfrac}
and the {\ttfamily scale} configuration parameters of
\multidrizzle. For the purpose of this paper, we assume $F_A$ to be
constant and equal to $F_A \sim 0.316$ ($p=0.8$ and $s=0.6$). By
assuming a constant corrective factor regardless of size, we make less
than a 10\% error on the noise estimation of the smallest objects
detected.

\begin{figure}[htb]
\epsscale{1.0} 
\plotone{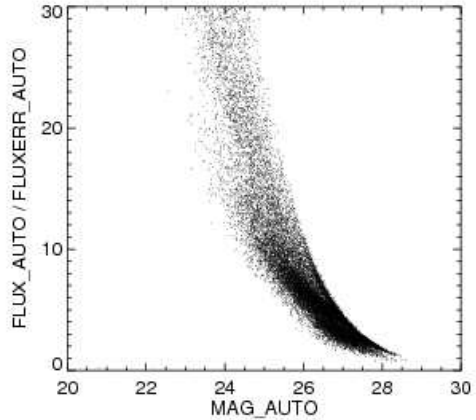}
\caption{ The significance of COSMOS detections defined as
  \textsc{flux\_auto}/\textsc{fluxerr\_auto} where
  \textsc{fluxerr\_auto} has been corrected for correlated
  noise.}
\label{sn}
\end{figure}

As implemented by SExtractor, the formulas for the flux and magnitude
uncertainties (for both \textsc{auto} and the \textsc{iso} quantities)
are given by:

\begin{equation}
\textsc{flux\_err} = \sqrt{A\sigma^2 +F/g}
\end{equation}

\begin{equation}
\textsc{mag\_err} = \frac{2.5}{\ln 10}\frac{\textsc{flux\_err}}{F}
\end{equation}

\noindent where $A$ is the area (in pixels) over which the flux $F$
(in ADU) is summed and $g$ is the detector gain. To correct the
magnitude and flux errors reported by SExtractor for the correlated
noise, we replace $\sigma$, the standard deviation of the noise (in
ADU) estimated by SExtractor, by $\sigma/\sqrt{F_A}$ within the above
equations. The significance of a COSMOS detection, after this
correction is applied, is defined as
$\textsc{s/n}=\textsc{flux\_auto}/\textsc{fluxerr\_auto}$ (see Figure
\ref{sn}).


\section{Quality Assessment of the ACS Catalog}\label{quality}

\subsection{Galaxy and Stellar Counts}\label{galcounts}

\begin{figure}[htb] 
\plotone{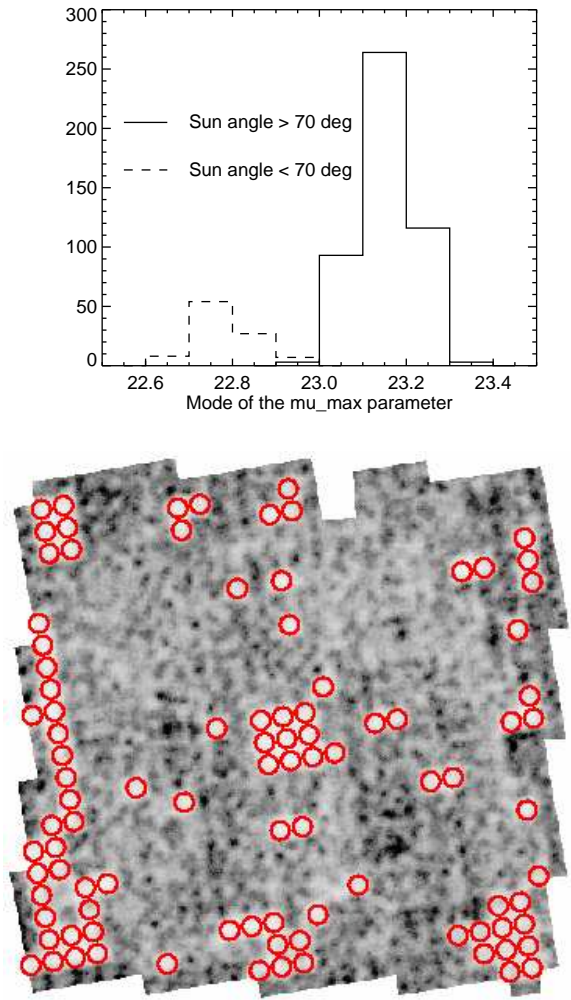}
\caption{ The top panel shows the histogram of $m_i$ (mode of the
  \textsc{mu\_max} parameter for each image) demonstrating that images
  with a sun angle less than 70 degrees are not quite as deep as
  images with a sun angle greater than 70 degrees. The bottom panel
  shows the density of faint objects ($27 < F814W < 28$) within the
  COSMOS field.  The red circles indicate the pointings for which the
  angle with the sun is less than 70 degrees.}
\label{sunmode}
\end{figure}

The same \textsc{mu\_max} parameter used to classify stars and
galaxies can also be used as an indication of the background level and
the depth of the data. For each image {\itshape i} we calculate the
mode $m_i$ of the \textsc{mu\_max} parameter. We then divide the $m_i$
into two bins according to the angle of the telescope with the Sun at
the moment of the pointing. The histogram of $m_i$ (Figure
\ref{sunmode}) for these two bins reveals that the depth of the data
is bimodal depending on whether the angle of the sun is less or
greater than a critical angle of 70 deg. This is visually evident when
we inspect the density map of the very faint objects (Figure
\ref{sunmode}). 96 pointings out of 575 have a Sun angle less than the
critical value making them slightly shallower than the average.

 
\begin{figure}[htb]
\plotone{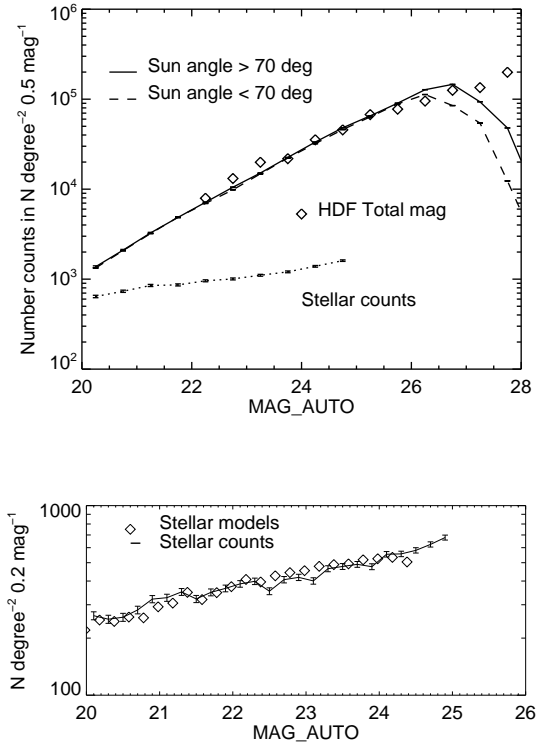}
\caption{ The top panel shows the galaxy and stellar number counts as
  compared to the HDF. The dashed curve corresponds to images with a
  sun-angle of less than 70 degrees and the solid curve corresponds to
  images with a sun-angle greater than 70 degrees. Poisson error bars
  are also indicated but are very small. The bottom panel shows the
  point source selection for the catalog compared to stellar models
  computed from evolutionary tracks and constrained by local Hipparcos
  data.}
\label{numcounts}
\end{figure}

The number counts serve as a check of the approximate photometric
calibration and the depth of the data. Figure \ref{numcounts} shows
the counts for galaxies and stars compared to the reported HDF F814W
counts \citep{Williams:1996}. The magnitudes are given in the
AB-system.  Stars have been subtracted from the galaxy counts up to
$F814W=25$.  At fainter magnitudes their contribution is
negligible.  We plot raw number counts only, {\it i.e.} we do not
correct for incompleteness at the faint end. To facilitate comparisons
with other surveys, we fit the galaxy counts between $F814W=20$
and $F814W=26$, to an exponential of the form $N=B \times 10^{A
  \times mag}$ where N has units of number degree$^{-2}$ 0.5
mag$^{-1}$. For the deeper set of images (sun angle $>$ 70 degrees),
we find $A=0.332$ and $log_{10}(B)=-3.543$. The raw galaxy and stellar
number counts are provided in Table \ref{counts}.

We also fit the stellar counts to models as shown in Figure
\ref{numcounts}. The star count predictions have been done using the
Besan{\c c}on model of the Galaxy \citep[2004]{Robin:2003} and are
described in detail in \citep{Robin:2007}. By extrapolating the
stellar number counts, we estimate that the galaxy catalog has less
than a 3\% contamination from stars at magnitudes greater than 25. The
fit to stellar models is excellent between $F814W=20$ and $F814W=25$,
and at magnitudes less than 19, we visually inspect the catalog to
check that the star selection is correct to within 0.5\%. This small
error arises mainly from false detections by SExtractor of the
diffraction spikes of bright, saturated stars.





\begin{deluxetable}{lcccc}
\tabletypesize{\scriptsize}
\tablecolumns{3}
\tablecaption{Cosmos F814W galaxy and stellar number counts \label{counts}}
\tablewidth{0pt} 
\tablehead{ 
 \colhead{$F814W$} & \colhead{Galaxy density} & \colhead{Stellar density}\\
 \colhead{  } & \colhead{$log_{10}(n)$ deg$^{-2}$ 0.5 mag$^{-1}$} & \colhead{$log_{10}(n)$ deg$^{-2}$ 0.5 mag$^{-1}$}
}
 \startdata 
20.25 & 3.138 $\pm$ 1.503  & 2.807 $\pm$ 1.297  \\ 
20.75 & 3.323 $\pm$ 1.600  & 2.865 $\pm$ 1.329  \\ 
21.25 & 3.514 $\pm$ 1.691  & 2.930 $\pm$ 1.361  \\ 
21.75 & 3.686 $\pm$ 1.777  & 2.936 $\pm$ 1.365  \\ 
22.25 & 3.853 $\pm$ 1.860  & 2.981 $\pm$ 1.385  \\ 
22.75 & 4.022 $\pm$ 1.945  & 3.003 $\pm$ 1.401  \\ 
23.25 & 4.180 $\pm$ 2.023  & 3.043 $\pm$ 1.414  \\ 
23.75 & 4.352 $\pm$ 2.110  & 3.081 $\pm$ 1.437  \\ 
24.25 & 4.523 $\pm$ 2.196  & 3.143 $\pm$ 1.463  \\ 
24.75 & 4.682 $\pm$ 2.275  & 3.206 $\pm$ 1.493  \\ 
25.25 & 4.814 $\pm$ 2.340  & \nodata  \\ 
25.75 & 4.956 $\pm$ 2.412  & \nodata  \\ \enddata
\tablecomments{Galaxy counts are derived for the 479 images with a
sunangle greater than 70 degrees. Magnitudes are the SExtractor \textsc{mag\_auto}}.
\end{deluxetable}

\subsection{Completeness}

\begin{figure}[htb] 
\plotone{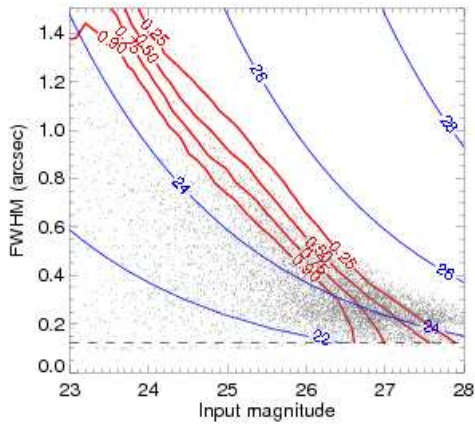}
\caption{ Completeness of the COSMOS F814W catalog as a function of
  total magnitude and FWHM determined by inserting fake objects into
  an ACS image. The thick contours show the percentage of fake objects
  recovered by SExtractor. The thin contours are the lines of constant
  surface brightness, in units of mag arcsec$^{-2}$, assuming a
  Gaussian profile. The grey points represent a random sample of
  objects from the COSMOS catalog plotted as a function of
  \textsc{mag\_auto} and \textsc{fwhm\_image}. The dashed horizontal
  line indicates the size of the ACS PSF. Note that the simulations
  only consider objects with Gaussian profiles whereas in reality, the
  COSMOS objects exhibit a wide variety of profiles.}
\label{complete}
\end{figure}

The probability that a galaxy enters our catalog will depend on its
size and surface brightness profile. To quantify the completeness and
detection limits of our SExtractor configuration, we insert fake
objects with a Gaussian profile of varying FWHM and total magnitude
into empty regions of an ACS image and test how well these objects can
be recovered with our pipeline. Each artificial source was considered
to be correctly detected if its centroid was within 10 pixels and the
\textsc{mag\_auto} parameter was within 0.5 mag of the input value.
From this analysis, we determine our completeness as a function of
magnitude and FWHM and the results are shown in Figure
\ref{complete}. The completeness is about 90\% for objects with a FWHM
of 0.2\arcsec\ at F814W=26.6. These values should only be used as a
rough estimate however, as we do not actually model galaxies, but use
a simple Gaussian profile for artificial objects.

 
\subsection{The COSMOS Redshift Distribution}\label{redshift distribution}

The estimation of the redshifts of galaxies is the major astrophysical
uncertainty inherent to weak lensing methods. To first approximation,
cosmic shear and tomography are mainly sensitive to the median
redshift of the sources while galaxy-galaxy lensing benefits greatly
from the knowledge of precise spectroscopic redshifts for the
foreground lenses \citep{Kleinheinrich:2005}.
With the depth and area coverage of COSMOS, we surpass the current
capability for complete spectroscopic follow-up. For most forthcoming
weak lensing surveys, this will also be the case, hence the importance
of the photometric redshift technique to measure redshifts for a
majority of the galaxies, to and beyond today's spectroscopic limits
\citep{Ilbert:2006}.
COSMOS presents a unique advantage, in terms of current weak lensing
surveys, of a prodigious multi-wavelength follow-up combined with the
planned measurement of $\sim 50 000$ spectroscopic redshifts by the
ongoing zCOSMOS program \citep{Lilly:2007}.  Upon completion, this
data set will provide the COSMOS lensing catalog with accurate
photometric redshifts and will be vital for refining and improving the
photometric technique in preparation for forthcoming weak lensing
surveys.
We present here a first analysis of the COSMOS redshift distribution.
A more detailed study of the systematic trends in the photometric
redshifts and of their effects on the redshift distribution is beyond
the scope of this paper and will be addressed elsewhere when more data
becomes available.
For the purposes of this paper, we adopt the magnitude dependent
parametrization of the redshift distribution, common to many other
weak lensing studies as given by \citet{Baugh:1993},

\begin{equation}
  \left\{
              \begin{array}{ll}
                   n(z,mag) \propto z^2 \exp [-\frac{z}{z_o(mag)}^{1.5}]\\
                   z_0(mag)=\frac{z_m(mag)}{1.412}
              \end{array}
       \right. 
\end{equation}

\noindent where $z_m$ is the median redshift of the survey as a function of
magnitude. We calculate $z_{m}$ for the COSMOS ACS data by bins of
$\Delta F814W=0.25$ for $20<F814W<24$ and derive the best
linear fit to $z_m$, given by:

\begin{equation}
z_{m}=(0.18\pm 0.01) \times F814W-(3.3\pm 0.2)
\end{equation}


The majority of the galaxies for which we have no redshift estimate at
$F814W<24$ are those in masked regions and their exclusion from this
derivation do not affect this results. At $F814W<24$, our redshift
incompleteness is less than $4\%$ and the dominant source of error is
the photometric redshift uncertainty expressed previously in Equation
\ref{photoz_err}.  We note that a significant number of galaxies
fainter then this limit do have photometric redshifts. The fact that
these galaxies represent a statistically incomplete sample, does not
matter for some applications. For example, these additional galaxies
are used in the cosmic shear measurement by \citet{Massey:2007}.

 
\begin{figure}[htb] 
\plotone{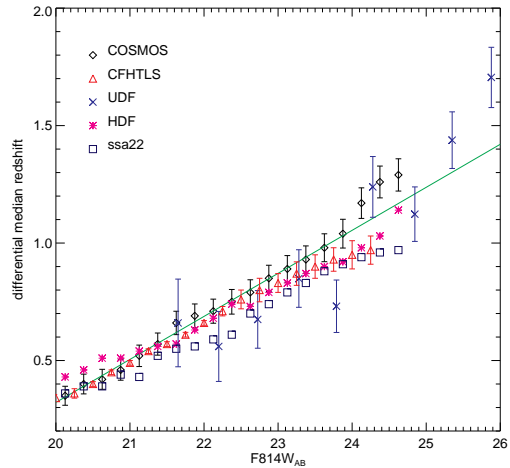}
\caption{Median redshift of COSMOS compared to various photometric
  redshift surveys. For clarity, error bars are only shown for the
  COSMOS, CFHTLS and UDF surveys}
\label{zm}
\end{figure}



\begin{deluxetable}{lcccc}
\tabletypesize{\scriptsize}
\tablecolumns{3}
\tablecaption{Present date photometric redshift surveys \label{compare surveys}}
\tablewidth{0pt}
\tablehead{
\colhead{Survey} & \colhead{Area}  & \colhead{Imaging data} & \colhead{Calibration spectra} & \colhead{Technique} } 
\startdata
COSMOS & 1.67 deg$^2$ & $B_j, V_j, g+, r+, i+, z+, NB816, u*, i*, Ks$ & 958 & COSMOS/BPZ \\
UDF & 11.97 arcmin$^2$ & $B, V, i', z', J$\tablenotemark{a} $H$\tablenotemark{b} & 76 & BPZ\\
CFHTLS & 3.2 deg$^2$ & $u*, g', r', i', z', J, K$ & 2867 & Le Phare\\
H-HDF-N & 0.2 deg$^2$ & $U_j, B_j, V_j, R_c, I_c, Z^+, HK'$ & 2149 & BPZ \\
SSA22 & 0.2 deg$^2$ & $U^*_j, B_j, V_j, R_c, I_c, z^+, J, H, K, HK'$ & 452 & BPZ \\
\enddata
\tablenotetext{a}{5.76 arcsec$^2$}
\tablenotetext{b}{160 arcmin$^2$}
\end{deluxetable}

In Figure \ref{zm}, we compare the median redshift of COSMOS to the
UDF survey \citep{Coe:2006}, the CFHTLS survey \citep{Ilbert:2006},
the H-HDF-N survey \citep{Capak:2004} and SSA22 (Hu et al. 2004, Capak
et al. 2004). 
Table \ref{compare surveys} is a summary of the data and
the methods employed by these different photometric redshift
surveys. All photometric redshifts have been computed with a Bayesian
prior based on luminosity functions. The agreement that we see between
the various surveys at $z<1$ is quite remarkable. At $z>1$ however,
the scatter in Figure \ref{zm} indicates the limits of current
photometric techniques.  Further simulations are clearly necessary in
order to understand the biases introduced in the redshift distribution
at $z>1$. Although we reserve a full discussion for a future paper, we
can already highlight some of the issues at hand.

The photometric redshift technique relies on detecting and measuring
the strength of broad spectral features.  These same features are used
by color selection techniques to select objects at specific redshifts.
The key features are the 4000\AA\ break, the Lyman break at 912\AA\~,
Lyman absorption at 1216\AA\~, and coronal line absorption between
1500-2500\AA\ (see Adelburger et. al. 2005).  Photometric redshifts
are very robust if one or more of these features are detectable in the
available data.  However, at faint magnitudes the photometric errors
and detection limits are often too large to constrain these features.

For example, problems arise for COSMOS between $1.5<z<3.2$ where the
measurable features, the 4000\AA\ break, and the coronal line
absorption features, are difficult to detect.  Indeed, at these
redshifts, the 4000\AA\ break is well into the IR where it is
difficult to obtain deep data.  A typical object at $z\simeq 2$ will
be $\sim1.4$ magnitudes fainter at $I$ than $K$.  This means objects
fainter than $F814W>23.5$ are not constrained by the present $K$-band
data.  At the other end of the spectrum the coronal absorption feature
has a typical strength of $\sim0.15$ magnitudes, which requires a
$25\sigma$ detection in $u^*$ to accurately differentiate from a
similar break in $z<0.5$ galaxies.  With the present $u^*$ data, this
corresponds to objects brighter than $F814W<24.5$ for typical
galaxies.
    
In conclusion, the high redshift tail of the COSMOS redshift
distribution will be more accurately determined with the forthcoming
NIR data and the future deep zCOSMOS spectroscopy which specifically
targets the $1.5<z<3$ region and will allow proper calibration down to
$F814W\sim 24$. A more detailed analysis of the photometric redshifts
will be conducted once the new data is available.

\section{PSF Correction and Shear Measurement}
\label{finalcat}

In this section, we measure the shapes of galaxies and correct them
for the convolution with the telescope's PSF and for other
instrumental effects. For each galaxy, we construct an unbiased local
estimator of the shear and derive the associated measurement error.

\subsection{PSF Modelling}\label{psf}

The ACS/WFC PSF is not as stable as one might naively hope from a
space-based camera. As shown in \citet{Rhodes:2007}, gradual changes
to both the size and the ellipticity pattern of the PSF due to
telescope ``breathing'', causes the PSF to change considerably on
timescales of weeks. The long period of time over which the COSMOS
field was observed forces us to take account of these variations (see
Figure \ref{fig:date_of_observation}). Although other strategies have
been demonstrated successfully for observations conducted on a shorter
time span, it would be inappropriate for us to assume, like
\citet{Lombardi:2005} or \citet{Jee:2005}, that the PSF is constant or
even, like \citet{Heymans:2005}, that the focus is piecewise constant.

\begin{figure}[htb]
\plotone{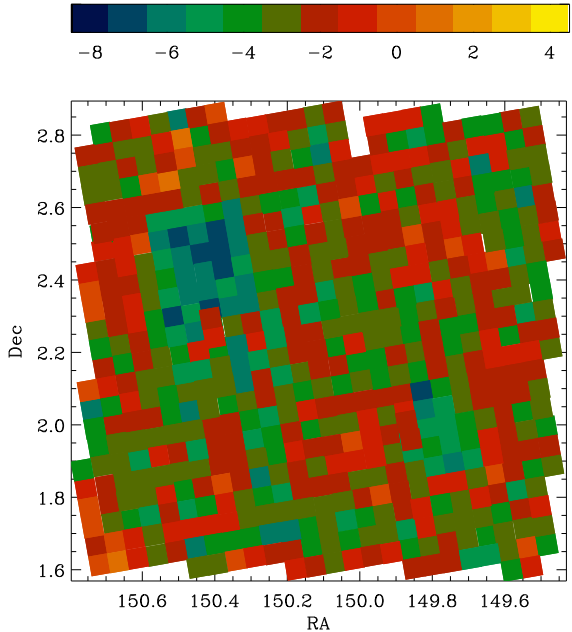}
\caption{Adopted PSF model across the survey. The colors correspond to
  deviations in the apparent focus of the telescope away from nominal
  ($\mu m$). These are caused by expansion and contraction of the HST
  due to thermal variations. Note that the focus values are clustered
  and not randomly distributed. See Rhodes et al. 2007 for more
  details about the PSF pattern at particular focus positions.}
\label{fig:focus_variation}
\end{figure}

Fortunately, most of the PSF variation can be ascribed to a single
physical parameter.  Thermal expansions and contractions of HST alter
the distance between the primary and secondary mirrors. As the
``effective focus'' deviates from nominal, the PSF becomes larger and
more elliptical, with the direction of elongation depending upon the
position above or below nominal focus \citep[\cf][]{Krist:2005}. The
thermal load on HST is constantly changing, in a complicated way, as
it passes in and out of the shadow of the Earth and is rotated to
different pointings.

As described in \citet{Rhodes:2007}, we have modified version 6.3 of
the {\sc TinyTim} ray-tracing program (Krist and Hook 2004) to create
a grid of model PSF images, at varying focus offsets. By comparing the
ellipticity of $\sim 20$ stars in each image to these, we can
determine the image's effective focus. Tests of this algorithm on ACS/WFC
images of dense stellar fields confirm that the best-fit effective
focus can be repeatably determined from a random sample of ten stars
brighter than $F814W=23$ with an rms error less than $1\mu$m. The
effective focus of the COSMOS images are shown in
Figure~\ref{fig:focus_variation}. An alternative correction scheme
based on PSF models constructed from dense stellar fields has also
been suggested by \citet{Schrabback:2006}.

Once images have been grouped by their effective focus position, we
can combine the few stars in each image into one large catalog. We
interpolate the PSF model parameters using a polynomial fit (of order
$3\times2\times2$ in each CCD separately), in the usual weak lensing
fashion \citep[\cf][]{Massey:2002}. See \citet{Rhodes:2007} for more
details concerning the PSF modelling scheme.


\subsection{Galaxy Shape Measurement}\label{shape}

We use the shape measurement method developed for space-based imaging
by \citet[][hereafter RRG]{Rhodes:2000}. The RRG method has been
optimized for space-based images with small PSFs and has previously
been used on weak lensing analyses of WFPC2 and STIS data
\citep{Rhodes:2001, Refregier:2002, Rhodes:2004}. In a manner similar
to the common ``KSB'' method \citep{Kaiser:1995}, RRG measures the
second and fourth order Gaussian-weighted moments of each galaxy:

\begin{equation}
\label{eq:moments}
I_{ij}=\frac{\sum w I x_i x_j }{\sum w I},
\end{equation}

\begin{equation}
\label{eq:moments4}
I_{ijkl}=\frac{\sum w I x_i x_j x_k x_l}{\sum w I}.
\end{equation}

\noindent The sum is over all pixels, $w$ is the size of the Gaussian
weight function, $I$ is the pixel intensity, and the $x_i$ coordinates
are measured in pixels. The Gaussian weight function is necessary to
suppress divergent sky noise contributions in the measurement of the
quadripole moments. The RRG method is well-suited to the small,
diffraction-limited PSF obtained from space, because it decreases the
noise on the shear estimators by correcting each moment for the PSF
linearly, and only dividing them to form an ellipticity at the last
possible moment.

After the moments have been corrected for the PSF, an ellipticity
$\varepsilon=(e_{1},e_{2})$ and size measure, $d$, are calculated for
each galaxy:

\begin{equation}
\label{eq:e1}
e_{1}=\frac{I_{xx}-I_{yy} }{I_{xx}+I_{yy}},
\end{equation}

\begin{equation}
\label{eq:e2}
e_2=\frac{2I_{xy}}{I_{xx}+I_{yy}},
\end{equation}

\begin{equation}
d=\sqrt{\frac{(I_{xx}+I_{yy})}{2}}.
\end{equation}

Note that the $d$ parameter is a measure of galaxy size but that its
value will depend on the choice of the width of the Gaussian weight
function, $w$.

\subsection{Shear Measurement}

The estimator $\varepsilon=(e_{1},e_{2})$ is not yet a shear
estimator, because it does not respond linearly to changes in
shear. It must first be normalized by a shear susceptibility factor
(also known as the ``shear polarizability''),

\begin{equation}
\tilde\gamma=\{\tilde\gamma_1,\tilde\gamma_2\} =\frac{\varepsilon}{G} ~,
\end{equation}

%

\noindent where the shear susceptibility factor, $G$, is measured from
moments of the global distribution of $\varepsilon$ and other, higher
order shape parameters \citep[see equation 28 in][]{Rhodes:2000}. The
RRG formalism does not allow for $G$ to be calculated for any
individual galaxy. However, $G$ can be calculated for an ensemble of
galaxies by averaging over a population's shape moments. Previous
incarnations of RRG that have been used to measure cosmic shear
\citep{Rhodes:2001, Refregier:2002, Rhodes:2004}, have made use of a
single value of $G$ for the entire survey. Adopting this approach for
the COSMOS data would yield a value of $G=1.13$. However, the Shear
TEsting Program \citep{Massey:2007} showed that $G$ can vary
significantly as a function of object flux, whether this be due to
evolution in galaxy morphologies as a function of redshift, or noise
in the wings of faint galaxies that simply impedes the measurement of
their radial profiles and higher order moments. An increase in shear
susceptibility with object $\sn$ has also been seen in KSB-type
analyses \citep{Massey:2004}.

\begin{figure}[htb]
\plotone{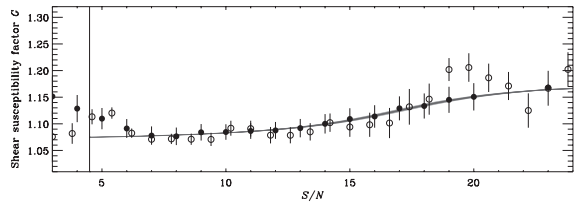}
\caption{Interpolation of the shear susceptibility factor $G$. The
  solid circles show $G$, calculated in bins of $\sn$, for the COSMOS
  data. The open circles show the same for the simulated COSMOS
  images. The solid line shows the shear susceptibility model adopted
  for the data.}
\label{fig:fitted_G}
\end{figure}

Our tests have confirmed that a constant value would be insufficiently
precise for a survey the size of COSMOS, and would particularly affect
the kind of 3D analysis for which COSMOS is so well-suited. We
therefore calculate $G$ from the COSMOS data in bins of $\sn$ (see
Figure~\ref{fig:fitted_G}). Simulated images of the COSMOS data are
created using the shapelets-based method of \citet{Massey:2004a} (see
$\S$\ref{simu}). The variations of $G$ as a function of $\sn$ are
apparent in the COSMOS data are well reproduced by the simulated
data. For COSMOS galaxies, we find that variations in $G$ as a
function of $\sn$ are well-fit by

\begin{equation}
\label{g}
G = 1.125+0.04\arctan\left(\frac{\sn-17}{4}\right).
\end{equation}

Adopting the above model, we derive $G$ for each galaxy as a function
of $\sn$.

 
\subsection{Calibration via Simulated Images}\label{simu}

Using the shapelets-based method of \citet{Massey:2004a}, we have
created simulated images with the same depth, noise properties, PSF,
and galaxy morphology distribution as the real COSMOS data. A known
shear signal was applied to the images, which we have then attempted
to measure using the same pipeline as the data. This exercise is
similar to the Shear TEsting Program
\citep[STEP;][]{Heymans:2006a,Massey:2006} but tailored exclusively to
COSMOS.

The simulated COSMOS images are each $4\arcmin\times 4\arcmin$, and
contain $\sim500$ galaxies after applying the same catalog cuts that
were applied to the real data (see $\S$\ref{cuts}). Simulated galaxy
morphologies are based on those observed in the Hubble Deep Fields
\citep[][]{Williams:1996,Williams:1998}, parametrized as shapelets and
randomly rotated/flipped before being sheared. Different input
galaxies were used in each simulated image, as if they were pointing
to different patches of the sky, to keep them independent. To simplify
later analysis, all of the galaxies within an image were sheared by
the same amount. A total of 41 images were made, with shears applied
in integer steps from $-10\%$ to $+10\%$ in the $\gamma_1$ component
(while $\gamma_2$ was fixed at zero) and similarly for the $\gamma_2$
component. The images were then convolved with a model ACS PSF. Again
to simplify the analysis, this was a constant PSF obtained from {\sc
  TinyTim}. Its $(e_1,e_2)$ ellipticity is
$(-0.21\%,-2.07\%)\pm(0.14,0.10)$. No stars were included in the
simulated images; a separate star field was created, from which the
PSF moments could be measured. Noise was added to all of these images,
to the same depth as the COSMOS observations, and with a similar (but
isotropic) correlation between adjacent pixels to mimic the effects of
\multidrizzle\ and unresolved background sources.

 
\begin{figure}[htb]
\plotone{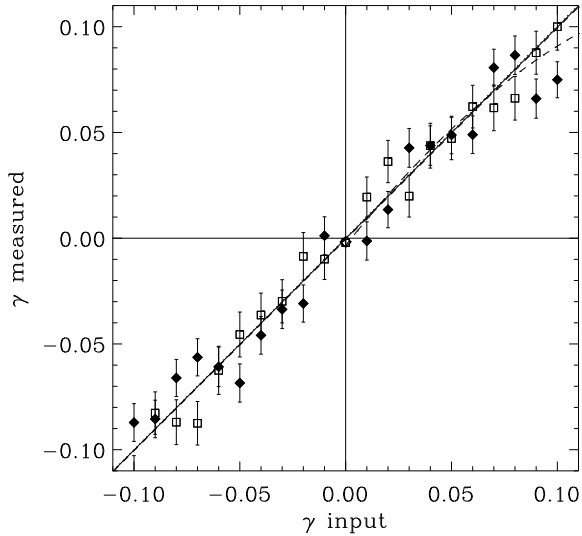}
\caption{ Calibration of the RRG shear measurement method from
  simulated COSMOS images containing a known input shear. The measured
  shear on the y-axis includes the shear calibration factor $C$.
  Squares show measurements of $\gamma_1$, diamonds show measurements
  of $\gamma_2$. The solid line is a linear fit to deviations from the
  ideal case of $\gamma_{\rm measured}=\gamma_{\rm input}$ for all
  points. The dashed line is a quadratic fit demonstrating that the
  curvature terms are negligible.}
\label{fig:simages}
\end{figure}

The recovered shear measurement from the simulated data is presented
in Figure~\ref{fig:simages}. We find that, in order to correctly
measure the input shear on COSMOS-like images, the RRG method requires
an overall calibration factor of $C=(0.86^{+0.07}_{-0.05})^{-1}$, so that

\begin{equation}
\tilde\gamma=C \times \frac{\varepsilon}{G} ~.
\end{equation}

The necessity for such calibration factors has long been known in the
field \citep[\eg][]{Bacon:2001,Erben:2001}, and is in accord with
results from STEP1. STEP2 and \citet{High:2007} suggest that this may
be intrinsic with KSB-related methods and, furthermore, that the
calibration can vary for the two components of shear.  For this
reason, we fit each component separately, and use final calibration
factors of $C_1=(0.80)^{-1}$ for $\tilde\gamma_1$ and
$C_2=(0.92)^{-1}$ for $\tilde\gamma_2$.
After this recalibration, Table~\ref{table:step} shows STEP-like
estimates of the additive bias $\langle c\rangle$ and the
multiplicative bias $\langle m\rangle$ obtained by fitting deviations
of the recovered shear from the input shear. Both of these are
consistent with ideal shear recovery, although the error on these
estimates will be propagated through subsequent analyses.

\begin{table}[tb]
\begin{center}
\vspace{3mm}
\begin{tabular}{cr@{~$\pm$~}l}
\tableline\tableline
\multicolumn{3}{l}{Linear STEP fit} \\
\tableline
$\langle c\rangle$   &  (~2.1 & $40.0)\times10^{-4}$ \\
~$c_1$               &  (~5.6 & $28.5)\times10^{-4}$ \\
~$c_2$               &  (-1.3 & $28.0)\times10^{-4}$ \\
$\langle m\rangle$   &  (-0.3 & $9.4)\times10^{-2}$ \\
~$m_1$               & (-12.2 & $6.5)\times10^{-2}$ \\
~$m_2$               & (~11.6 & $6.3)\times10^{-2}$ \\
\tableline
\hspace{22mm}\vspace{3mm} \\
\tableline\tableline
\multicolumn{3}{l}{Quadratic STEP fit} \\
\tableline
$\langle c\rangle$   & (-23.3 & $121.8)\times10^{-4}$ \\
$\langle m\rangle$   & (~20.1 & $51.5)\times10^{-2}$ \\
$\langle q\rangle$   &  -2.84 &  4.59 \\
\tableline
\end{tabular}

\caption{--- Calibration of the RRG shear measurement method on simulated COSMOS
images containing a known shear, described using STEP parameters. Figures are
supplied after the application of the shear calibration
factor.\label{table:step}}

\end{center}
\end{table}

For the kind of 3D shear analysis for which COSMOS is so well-suited,
the simultaneous calibration of shears from an entire population of
galaxies is insufficient. Since our companion paper,
\citet{Massey:2007}, is concerned with the growth of the shear signal
as a function of redshift, it is crucial that the shear calibration be
equally precise for both distant and relatively nearby galaxies. Given
that more distant galaxies are fainter and smaller and that the
details of the shear measurement depend upon a fixed PSF size, pixel
size, and noise level, this requirement is not trivial. We have
therefore split the simulated galaxy catalog in half by magnitude (at
$F814W=25.4$) and by size (at $d=5.0$ pixels), and repeated the
analysis. We find that our shear calibration $\langle m\rangle$ is
robust for galaxies of different fluxes within $1\%$ and of different
sizes within $4\%$. Redshifts were not available for the simulated
galaxies, so a direct split in redshift was not possible. Although
this effect is clearly small in the regime of our current
measurements, it will be significant in future weak lensing surveys
where the error budget will be dominated by systematic uncertainties.




\subsection{Error on the Shear Estimator}\label{error_shear}

For each galaxy, the error on the measured shear is estimated using
the same method as implemented in the {\sc Photo} pipeline
\citep{Lupton:2001} to analyze data from the Sloan Digital Sky
Survey. For each object, we assume that the optimal moments are the
same as the moments corresponding to a best fit Gaussian. This
formulation allows us to determine the covariance matrix of the
moments in the usual way of non-linear least squares. To be precise,
we model the image with a 2D elliptical Gaussian model
\begin{eqnarray}
M({\bf x})  =  \frac{f}{2 \pi |Q|^{1/2}} \times \nonumber \\
  \exp[ - \frac{1}{2} ({\bf x-\mu})^{T} Q^{-1} {\bf (x-\mu)}].
\end{eqnarray}
This model has six parameters: the flux, $f$, the two centroids, ${\bf
  \mu}$, and the three moments which form the elements of the
symmetric matrix $Q$. These parameters are noted ${p_l}$. Next, we
derive the $\chi^2$ in the usual way

\begin{equation}
  \chi^2 \equiv \frac{1}{\sigma^2} \sum_{ij} [M({\bf x_{ij}}) -I_{ij}]^2,
\end{equation}
where $\sigma$ is the sky noise level.  We can now compute the
$6\times6$ Fisher matrix which is the matrix of the second derivatives
of the $\chi^2$
\begin{eqnarray}
F_{kl} & \equiv & \frac{1}{2} \frac{\partial^2 \chi^2}{\partial p_k \partial p_l} \nonumber \\
& = & \frac{1}{\sigma^2} \sum_{ij} \frac{\partial M({\bf x_{ij}})}{\partial p_k} \frac{\partial M({\bf x_{ij}})}{\partial p_l} \nonumber \\
& - & \frac{1}{\sigma^2} \sum_{ij} [M({\bf x_{ij}}) -I_{ij}] ~\frac{\partial^2 M({\bf x_{ij}})}{\partial p_k \partial p_l}. \nonumber
\end{eqnarray}
As is customary, we drop the second term which is proportional to the
residuals. This term is usually very small compared to the first.  The
covariance matrix of the Gaussian parameters are then the inverse of
the Fisher matrix. The $3\times3$ block of the covariance matrix
corresponding to the second moments can then be extracted. Because the
whole $6\times6$ matrix was inverted, this correctly marginalizes over
centroid errors and other model parameter degeneracies. In this way,
it differs from formulas which assume a constant (non-adaptive)
weighting function and perfect centroiding.

This Fisher matrix does not depend explicitly on the data; it only
depends on the best fit parameters and can be computed analytically.
It is simply a function of four numbers: the three second order
moments defined in Equation~\ref{eq:moments} and the signal-to-noise
ratio ($f/\sigma$), which can be parametrized by the magnitude error
(for example, \textsc{magerr\_auto}). Since the flux computed with
SExtractor is not exactly equal to $f$ (which would be the best fit
Gaussian amplitude) we allow for a single calibration factor and
multiply the covariance matrix by this.  We calibrate this factor with
image simulations and verify that the errors are correctly predicted.

Since the ellipticity components are computed from the moments, the
variances of the ellipticity components can be computed by linearly
propagating the covariance matrix of the moments. Finally, the two
ellipticity components can be shown to be uncorrelated with each
other.



\section{Final Galaxy Selection}\label{cuts}

\subsection{Lensing Cuts}

Estimations of the gravitational shear will be improved by averaging
only those galaxies with precise shape measurements on the condition
that no ellipticity selection bias is introduced by the ``lensing
cuts''. We apply strict cuts to the $\mathcal{C}_2$ catalog that are
designed to extract a sample of resolved galaxies with reliable shape
measurements. The resulting catalog is referred to as
$\mathcal{C}_4$. Our ``lensing cuts'' are summarized in Table
\ref{lenscuts} and are based on the four following parameters:

\begin{enumerate}
\item The estimated significance of each galaxy detection, where the
  significance is defined as
  $\textsc{s/n}=\textsc{flux\_auto}/\textsc{fluxerr\_auto}$,
\item The first order moments, $I_{xx}$ and $I_{yy}$,
\item The total ellipticity, $e=\sqrt{e_{1}^2+e_{2}^2}$,
\item The galaxy size as defined by the RRG $d$ parameter (see \S \ref{shape}).
\end{enumerate}



\begin{deluxetable}{lc}
\tabletypesize{\scriptsize}
\tablecolumns{2}
\tablecaption{Lensing cuts applied to $\mathcal{C}_2$\label{lenscuts}}
\tablewidth{0pt}
\tablehead{
\colhead{Parameter}  & \colhead{Galaxies retained in $\mathcal{C}_4$}} 
\startdata 
$I_{xx}$ and $I_{yy}$  & Finite\tablenotemark{a} $I_{xx}$ and $I_{yy}$ \\
RRG size parameter\tablenotemark{b} $d$ &  $d > 3.6$ pixels\\
Significance & $S/N > 4.5$   \\
Total ellipticity\tablenotemark{c} $e$ & $e < 2$ \\
\enddata
\tablenotetext{a}{Indicating that the RRG code converged}
\tablenotetext{b}{Uncorrected for the PSF}
\tablenotetext{c}{Corrected for the PSF}
\end{deluxetable}

The final size cut is designed to select galaxies with well resolved
shapes. Indeed, PSF corrections become increasingly significant as the
size of a galaxy approaches that of the PSF and the intrinsic shape of
a galaxy becomes more difficult to measure.  In COSMOS images, the
typical size (as defined by $d$) of a star is about $d\star=2.2$
pixels (0.066\arcsec). Our size cut is thus equivalent to selecting
galaxies with $d_g>1.6 \times d\star$. Note that in this section, as
well as in all following sections, $d$ has not been corrected for the
PSF.


The ellipticity cut at $e<2$ may be surprising given that, by
definition, ellipticities are restricted to $e \leq 1$. In reality,
however, because of noise, it is possible to measure an ellipticity of
$e>1$. Selecting galaxies with $e<1$ could introduce an unwanted
ellipticity bias, but, because we are only interested in ensemble
averages, an acceptable solution is to cut out only a small number of
large outliers ($e>2$).

Note that further cuts may be required for some applications (for
example, to isolate only those objects with well-measured photometric
redshifts). In addition to these cuts, a galaxy-by-galaxy weighting
scheme may also be used to minimize the impact of shape measurement
noise.

\subsection{Effective Galaxy Number Density}

Once the ``lensing cuts'' have been applied, the $\mathcal{C}_4$
catalog contains only those galaxies which are useful for lensing
analyses with the COSMOS data. Predictions for future weak lensing
surveys based on COSMOS results must consider the ``effective number''
of galaxies that are actually useful for lensing purposes (and not the
number of raw detections). With these considerations in mind, we
define the ``effective galaxy number density'', $N_{g}(z)$, as the
total number of galaxies within $\mathcal{C}_4$, per unit area, and
with redshifts below a given redshift, $z$. The equivalent quantities
as a function of galaxy magnitude and size are $N_{g}(m)$ and
$N_{g}(d)$. For each galaxy, the SExtractor \textsc{mag\_auto}
parameter is used to estimate the magnitude and the RRG $d$ parameter
is used as an estimate of the size. We also consider the derivatives
of $N_{g}(z)$, $N_{g}(m)$, and $N_{g}(d)$ so that:

\begin{equation}\label{neff}
  N_{g}(z) = \int_{0}^{z} n_{g}(z')dz', 
\end{equation}
\begin{equation}
  N_{g}(m) = \int_{20}^{m} n_{g}(m')dm',
\end{equation}
\begin{equation}
  N_{g}(d) = \int_{d}^{\infty} n_{g}(d')dd'.
\end{equation}

\begin{figure}[htb]
\epsscale{0.9}
\plotone{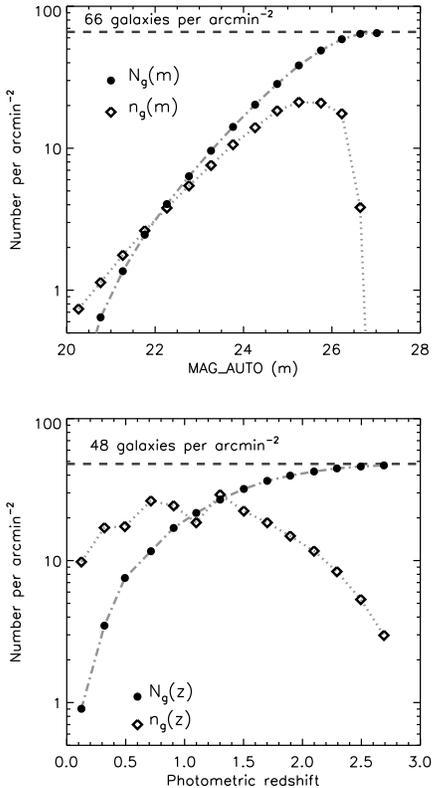}
\caption{ In the upper panel we show the effective number density of
  galaxies as a function of $F814W$ magnitude. Only resolved galaxies
  with precise shape measurement are included in these counts. In the
  lower panel, we show the effective number density of galaxies as a
  function of redshift. The effective density still evolves sharply
  after $z>1$ demonstrating that the COSMOS lensing data is a powerful
  probe of structures at $z<1$.  }
\label{lenscounts}
\end{figure}

\begin{figure}[htb] 
\epsscale{0.9}
\plotone{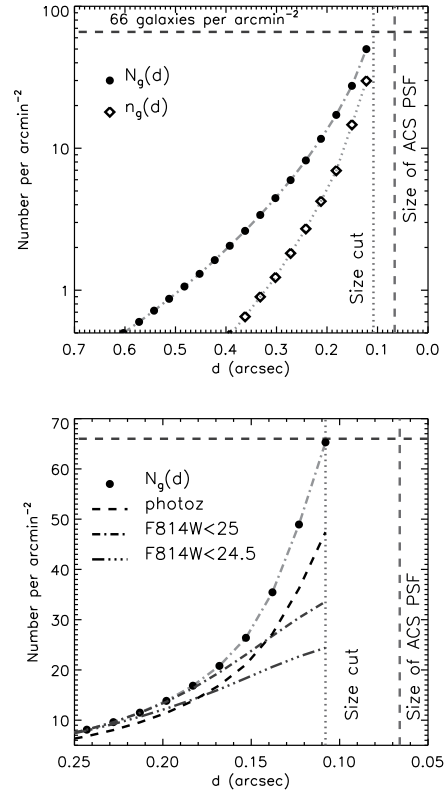}
\caption{ In the upper panel we show the effective number density of
  galaxies as a function of size. The dotted vertical line indicates
  the size cut that we make in order to extract galaxies with precise
  shape measurements. This size cut is such that $d_g>1.6 \times
  d\star$. $N_{g}$ rises very sharply as a function of decreasing $d$
  demonstrating that small galaxies make up the majority of our
  lensing sources. The lower panel shows $N_{g}$ for COSMOS galaxies
  with accurate photometric redshifts and for magnitude cuts of
  $F814W<25$ and $F814W<24.5$.  }
\label{lenscounts2}
\end{figure}

The total number density of galaxies in $\mathcal{C}_4$ is noted as
$N_{g}$ and is significantly lower than the number density of {\it
  detected} galaxies. In total, the final lensing catalog
$\mathcal{C}_4$ contains 3.9$\times 10^5$ galaxies with accurate shape
measurements and 2.8$\times 10^5$ galaxies with both shape and
photometric redshift measurements. Table \ref{final_table} shows a
summary of the different steps leading to this catalog. The surveyed
area of COSMOS is 1.64 deg$^2$ leading to an overall number density of
$N_{g}\sim$66 galaxies per arcmin$^2$ for the first sample and
$\sim$50 for the second. These numbers can be contrasted to a more
sparse 15-25 galaxies per arcmin$^2$ typically resolved with deep,
ground-based surveys. In Figures~\ref{lenscounts} and
\ref{lenscounts2} we show the effective densities defined above as
well as their corresponding derivatives. From these figures, we can
draw the following conclusions:

\begin{itemize}
\item Over $60\%$ of the COSMOS source galaxies are at redshifts
  higher than $z=1$. The COSMOS weak lensing data is therefore a
  powerful probe of the dark matter distribution from $z \sim 1$ to
  the present day.
\item About $18$ galaxies per arcmin$^{-2}$ (73\%) are discarded when
  we select only those galaxies with accurate photometric
  redshifts. The primary cause of this loss are the larger areas
  masked out in the ground-based data as compared to the ACS
  data. Future space-based weak lensing surveys could recover the
  remaining 27\% by using space-based, multi-wavelength imaging to
  derive photometric redshifts.
\item The effective number density rises very steeply with decreasing
  galaxy size; over $50\%$ of our total number of sources have $d<5$
  pixels ($0.15$\arcsec). The current size cut for the COSMOS data is
  $d=3.6$ pixels and the size of the ACS PSF is $d\star=2.2$ pixels
  (0.066\arcsec). By pushing this size barrier to even smaller values,
  future surveys could very quickly obtain much higher effective
  densities.
\end{itemize}

Finally, the cuts that we have applied to the COSMOS data (in
particular the size cut) are stricter than would be necessary without
significant CTE effects that degrade the shape measurements of the
faintest galaxies. Next generation space-based missions designed to
avoid the problems encountered and identified in COSMOS, as well as
future implementations of the COSMOS catalog, will undoubtedly achieve
higher effective number densities.

 
\section{Intrinsic Shape Noise: a Fundamental Limit to Weak Lensing
  Measurements}\label{shapenoise}

Under the assumption of weak gravitational lensing, a source galaxy
with intrinsic shape $\varepsilon_{int}$ and observed ellipticity
$\varepsilon_{obs}$ is related to the gravitational lensing induced
shear $\gamma$ according to:

\begin{equation}\label{eq:1}
  \varepsilon_{obs} = \varepsilon_{int}+\gamma.
\end{equation}

Throughout this paper, the gravitational shear is noted as $\gamma$
whereas $\tilde\gamma$ represents our estimator of $\gamma$. The above
relationship indicates that galaxies would be ideal tracers of the
distortions caused by gravitational lensing if the intrinsic shape
$\varepsilon_{int}$ of each source galaxy was known \textit{a
  priory}. A quick glance at an ACS image however, reveals that
galaxies display a very wide variety of shapes which unfortunately
prevents the extraction of $\gamma$ for any single galaxy.  Lensing
measurements thus exhibit an intrinsic limitation, encoded in the
width of the ellipticity distribution of the galaxy population, noted
here as $\sigma_{int}$, and often referred to as the ``intrinsic shape
noise''. Because the shape noise (of order $\sigma_{int} \sim 0.26$)
is significantly larger than weak shear (typically $\gamma \sim 0.01$
for cosmic shear), $\gamma$ must be estimated by averaging over a
large number of galaxies. In this case equation \ref{eq:1} simplifies
to:

\begin{equation}\label{eq:2}
<\varepsilon_{obs}> = <\gamma>.
\end{equation}
 
The uncertainty in the shear estimator, $\sigma_{\tilde\gamma}$,
arises from a combination of unavoidable intrinsic shape noise,
$\sigma_{int}^2=<\varepsilon_{int}^2>$ and the measurement error of
galaxy shapes $\sigma_{meas}^2$:

\begin{equation}\label{eq:3}
  \sigma_{\tilde\gamma}^2 = \sigma_{int}^2+\sigma_{meas}^2.
\end{equation}

In the following analysis, $\sigma_{\tilde\gamma}$ will be referred to
as the {\it shape noise} and $\sigma_{int}$ will be called the {\it
  intrinsic shape noise}. The former includes the shape measurement
error, $\sigma_{meas}$, and hence will vary according to the data-set
as well as the shape measurement method that is employed. Note that
the uncertainty contributions from, photon noise, PSF correction, CTE
calibration, and the shape measurement method are all included in our
definition of $\sigma_{meas}^2$. The weak lensing distortions averaged
over the whole COSMOS field are small, and represent a negligible
perturbation to equation \ref{eq:3}.

Instead of the simple arithmetic mean of equation \ref{eq:2}, many
lensing practitioners adopt in some form or another, an optimized
weighting scheme in order to estimate $\gamma$ which often
incorporates both the measurement error and the shape noise \citep[for
e.g.,][]{Bernstein:2002}. A constant value (of order $0.3$) is often
assumed for \shapenoise\ . However, it would not be surprising that
the same processes that shape galaxy formation also lead to a
variation of \shapenoise\ as a function of magnitude, galaxy type, or
redshift. Furthermore, as weak lensing surveys increase in both scale
and depth, it of intense interest to obtain accurate estimates of the
intrinsic shape noise floors that these surveys must confront. For
these reasons, we undertake a measurement of the shape noise,
$\sigma_{\tilde\gamma}$, as well as the intrinsic shape noise,
$\sigma_{int}$, as a function of magnitude, size and redshift. A more
detailed analysis of the intrinsic shape noise as a function of galaxy
morphology will be the subject of a future paper.

First, we estimate the shape noise $\sigma_{\tilde\gamma}$ directly
from the COSMOS data as a function of size and $F814W$ magnitude. To
derive $\sigma_{\tilde\gamma}$ we consider:

\begin{itemize}
\item The mean variance of both shear components (including correction
  factors), $\sigma_{\tilde\gamma}=(\sigma_{\tilde\gamma
    1}+\sigma_{\tilde\gamma 2}) / 2$
\item Galaxies with well measured shapes (see $\S$ \ref{cuts})
\end{itemize}


 
\begin{figure}
\epsscale{0.9}
\plotone{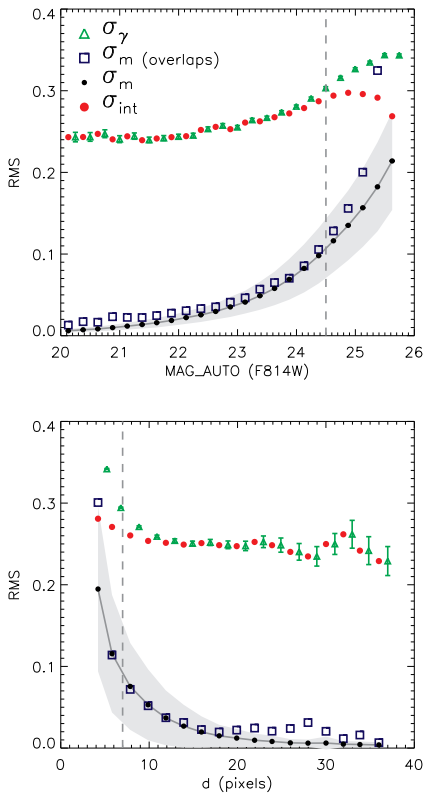}
\caption{ The observed scatter in the shear as a function of $F814W$
  magnitude and size. The scatter is a combination of intrinsic shape
  noise, $\sigma_{int}$, and a shape measurement error,
  $\sigma_{meas}$. The shape measurement error is determined by a
  theoretical model and tested using galaxies from overlapping
  regions. Shaded regions indicate the RMS width of the measurement
  error distribution. The intrinsic shape noise appears to increase
  slightly as a function of magnitude but is independent of size. At
  faint magnitudes and small galaxy sizes, the shear scatter increases
  rapidly due to large measurement errors. The dashed vertical line
  indicates the size and magnitude cut that we apply in order to
  construct Figure~\ref{sigma_z}.  }
\label{sigma_mag}
\end{figure}

\begin{figure}[h]
\epsscale{0.92}
\plotone{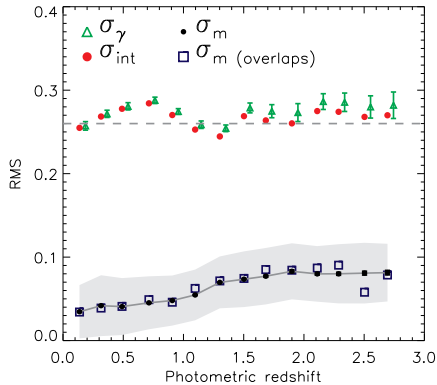}
\caption{ Intrinsic shape noise as a function of photometric
  redshift. Galaxies have first been selected to have measurement
  errors less than 0.1 ($d>7$ and $\textsc{mag\_auto}<24.5$) so that
  the scatter in the observed shear closely matches the intrinsic
  shape noise. A linear fit to the shape noise as a function of
  redshift reveals a flat distribution with a mean of
  $\sigma_{int}=0.26$.}
\label{sigma_z}
\end{figure}

Second, we derive an empirical estimation of the shape measurement
error, $\sigma_{meas}$, using a sample of $27 000$ galaxies that
belong to overlapping regions of adjacent pointings. Each of these
galaxies provide us with two independent shape measurements. Using
these overlaps, we find that the shape measurement error is a function
of both size and magnitude and increases beyond $\sigma_{meas} =0.1$
for $\textsc{mag\_auto}>24.5$ and $d<7$. Third, we compare this
empirical determination of $\sigma_{meas}$ to the theoretical one
derived in $\S$ \ref{error_shear}. We find that the theoretical model
of $\S$ \ref{error_shear} does remarkably well in predicting
$\sigma_{meas}$ as a function of size and magnitude. Thus confident in
the validity of this model, we adopt it for subsequent
derivations. Finally, using equation \ref{eq:3} and the shear
measurement error $\sigma_{meas}$, we extract the intrinsic shape
noise of our galaxy sample as a function of size, magnitude and
redshift. The results are shown in Figures~\ref{sigma_mag} and
\ref{sigma_z}.
 
As can be seen in Figure~\ref{sigma_mag}, large measurement errors
lead to an increase of $\sigma_{\tilde\gamma}$ at small sizes and
faint magnitudes. The intrinsic shape noise however, appears to change
little with either size of magnitude and remains constant at a value
of $\sigma_{int} \sim 0.26$. The slight apparent increase of the
intrinsic shape noise at fainter magnitudes is probably due to the
simplified measurement error estimator that we are using (indeed, the
overlaps indicate slightly higher errors). From this analysis, we can
draw the following conclusions:

\begin{itemize}
\item The intrinsic shape noise varies little from $z=0$ to
  $z=3$. Deep space based weak lensing surveys will therefore confront
  equivalent intrinsic shape noise floors as their shallower
  counterparts.

\item Measurement errors lead to an increase in the shape noise as a
  function of size and magnitude. A joint improvement in both imaging
  quality as well as shape measurement methodology will lead to shape
  noises that are closer to the intrinsic floor of 0.26.

\item We have yet to explore if the intrinsic shape noise varies as a
  function of galaxy type. If so, shear measurements could be improved
  by incorporating a galaxy-type discriminant into the weighting
  scheme. This will be the focus of a future paper.
\end{itemize}

\section{Conclusion}\label{conclusions}

We have carefully constructed a weak lensing catalog from 575 ACS/HST
tiles, the largest space-based survey to date. We have established the
quality of this catalog and analyzed the COSMOS redshift distribution,
showing broad agreement with other photometric redshift surveys out to
$z \sim 1$. The photometric redshifts are currently limited by the
lack of deep $K$-band data but will rapidly improve as this data soon
becomes available. Shapes have been measured for over $3.9 \times
10^5$ galaxies and corrected for distortions induced by PSF and CTE
effects. Simulations have been used in order to calibrate our shear
measurement method and a STEP-like analysis has been performed,
demonstrating our ability accurately measure shear with negligible
additive and multiplicative bias. The effective number density of the
COSMOS weak lensing catalog is $66$ galaxies per arcmin$^{-2}$ ($48$
when we consider only those with accurate photometric redshifts). A
large fraction of these galaxies are at $z>1$ making COSMOS a powerful
probe of the dark matter distribution and its evolution from $z=1$ to
the present day. The COSMOS survey is also of foremost importance for
the preparation and design of future wide field space-based lensing
missions. Regarding the design of such missions, our main conclusions
from working with the COSMOS data are the following:\\

\begin{enumerate}
\item Understanding and correcting for the time varying PSF and
  calibrating CTE effects were two of the most difficult challenges
  encountered with the COSMOS data. Reducing these two systematic
  effects should be a key specification in the design of
  next-generation telescopes and instruments.

\item Because 1) small galaxies are not as readily detected from the
  ground and 2) because larger areas are masked out in the
  ground-based data than in the ACS imaging, we lose 27\% of our
  source sample when we select only those with accurate photometric
  redshifts.

\item The effective number density of galaxies is a very sensitive
  function of survey depth and resolution. The capability to resolve
  and accurately measure the shapes of very \textit{small},
  \textit{faint} galaxies will be key in obtaining number densities of
  over 66 galaxies arcmin$^{-2}$.

\item Finally, we have derived the intrinsic shape noise of the galaxy
  sample and demonstrated that it remains fairly constant as a
  function of size, magnitude and redshift. Weak lensing measurements
  with deep space-based imaging are therefore on par with more shallow
  imaging in terms of the intrinsic shape noise floors that they must
  overcome.
\end{enumerate}

The COSMOS weak lensing data described in this paper has already been
used to measure cosmological parameters and to demonstrate the
feasiblity of the ``tomography'' technique
\citep{Massey:2007}. Striking weak lensing mass maps of the COSMOS
field have been made that reveal tantalizing evidence of a complex
interplay between the baryon and the dark matter distribution
\citep{Massey:2007a}. Yet more lensing analyses are underway including
a \textit{galaxy-galaxy lensing} and a \textit{group-galaxy lensing}
study that will undoubtedly lead to a better understanding of the
relationship between baryonic and dark matter structures and of its
evolution over cosmic time.
\clearpage

 

\acknowledgments
\noindent {\bf Acknowledgments}

The HST COSMOS Treasury program was supported through NASA grant
HST-GO-09822. We wish to thank our referee for useful comments and
Kevin Bundy for carefully reading the manuscript.  We also thank Tony
Roman, Denise Taylor, and David Soderblom for their assistance in
planning and scheduling of the extensive COSMOS observations.  We
gratefully acknowledge the contributions of the entire COSMOS
collaboration consisting of more than 70 scientists.  More information
on the COSMOS survey is
available \\
at {\bf \url{http://www.astro.caltech.edu/$\sim$cosmos}}. It is a
pleasure to acknowledge the excellent services provided by the NASA
IPAC/IRSA staff (Anastasia Laity, Anastasia Alexov, Bruce Berriman and
John Good) in providing online archive and server capabilities for the
COSMOS data-sets.  The COSMOS Science meeting in May 2005 was
supported in part by the NSF through grant OISE-0456439.  CH is
supported by a CITA National fellowship and, along with LVW,
acknowledges support from NSERC and CIAR. In France, the COSMOS
project is supported by CNES and the Programme National de
Cosmologie. JPK acknowledges support from CNRS.

\bibliographystyle{apj}


 \end{document}